\documentclass{IEEEtran}
\usepackage{cite}
\usepackage{amsmath,amssymb,amsfonts}
\usepackage{algorithmic}
\usepackage{graphicx}
\usepackage{subfigure}
\usepackage{threeparttable}
\usepackage{booktabs}
\usepackage{listings}
\usepackage{caption}

\newcommand*\phantomsubfigure[1]{\subfigure{\label{#1}}}
\begin{document}

\title{Autonomous Probabilistic Coprocessing with Petaflips per Second}
\author{Brian Sutton, Rafatul Faria, Lakshmi A. Ghantasala, Risi Jaiswal, Kerem Y. Camsari,
and Supriyo Datta

\thanks{BS, RF, LAG, RJ and SD are with the School of Electrical and Computer Engineering, Purdue University, West Lafayette, IN, 47907, USA, KYC is with the Department of Electrical and Computer Engineering, University of California, Santa Barbara, Santa Barbara, CA, 93106, USA. This work was supported in part by ASCENT, one of six centers in JUMP, a Semiconductor Research Corporation (SRC) program sponsored by DARPA and in part by the Center for Probabilistic Spin Logic for Low-Energy Boolean and Non-Boolean Computing (CAPSL), one of the Nanoelectronic Computing Research (nCORE) Centers, a Semiconductor Research Corporation (SRC) program sponsored by the NSF.}}



\maketitle
\begin{abstract}
In this paper we present a concrete design for a probabilistic (p-) computer based on a network of p-bits,  robust classical entities fluctuating between -1 and +1, with probabilities that are controlled through an input constructed from the outputs of other p-bits. The architecture of this probabilistic computer is similar to a stochastic neural network with the p-bit playing the role of a binary stochastic neuron, but with one key difference: there is no sequencer used to enforce an ordering of p-bit updates, as is typically required. Instead, we explore \textit{sequencerless} designs where all p-bits are allowed to flip autonomously and demonstrate that such designs can allow ultrafast operation unconstrained by available clock speeds without compromising the solution's fidelity. Based on experimental results from a hardware benchmark  of the autonomous design and benchmarked device models, we project that a nanomagnetic implementation can scale to achieve petaflips per second with millions of neurons. A key contribution of this paper is the focus on a hardware metric $-$ \textit{flips per second}$-$ as a problem and substrate-independent figure-of-merit for an emerging class of hardware annealers known as Ising Machines. Much like the shrinking feature sizes of transistors that have continually driven Moore's Law, we believe that flips per second can be continually improved in later technology generations of a wide class of probabilistic, domain specific hardware. 
\end{abstract}
 


\section{Introduction}
\label{sec:introduction}
\IEEEPARstart{S}{tochastic} artificial neural networks (ANN) have broad utility in optimization and machine learning (ML) tasks such as inference and learning\cite{schuman_survey_2017}. Even though stochastic ANNs are relatively rare in modern ML architectures \cite{schuman_survey_2017}, stochasticity is often viewed as a useful resource \cite{detorakis2019inherent,buesing2011neural,courbariaux2016binarized}. A class of emerging hardware accelerators, known as Ising Machines,  typically have stochastic neural network representations. Ising Machines designed to solve hard problems in combinatorial optimization continue to emerge using a wide-range of underlying technologies. Solvers for such problems have been explored using quantum effects, optical approaches, digital logic, and magnetic technologies \cite{boixo_evidence_2014, yamaoka_20k-spin_2016, mcmahon_fully-programmable_2016,inagaki_coherent_2016, okuyama_ising_2017,sutton_intrinsic_2017, hamerly_experimental_2019, wang_oscillator-based_2017, wang_oim:_2019, raychowdhury_computing_2019,aramon_physics_2019,goto_combinatorial_2019,yamamoto_statica_2020,su_cim_2020}. In general, these systems map a given optimization problem onto a hardware whose operation is guided by a cost function\cite{lucas_ising_2014,dattani2019quadratization}.

A common version of such stochastic neural networks is based on the concept of a binary stochastic neuron (BSN) \cite{ackley_learning_1985,neal_connectionist_1992} which fluctuates between -1 and +1 with probabilities that can be controlled through an input, $I_i$, constructed from the outputs of other BSNs, $m_j$. The synaptic function, $ I_i (\{m\})$, can have many different forms depending on the desired functionality,  but we will restrict our discussion to linear functions defined by a set of weights $W_{ij}$ such that
\begin{equation}
I_i (t+\tau_S) = \beta \ \sum_{j} W_{ij} m_j(t) 
\label{eq:synaptic_function}
\end{equation}
\noindent where $\beta$ is a constant and $\tau_{S}$ is the `synapse time', that is the time it takes to recompute the inputs $\{I\}$ every time the outputs $\{m\}$ change. In software implementations, each BSN is updated repeatedly according to
\begin{equation}
m_i (t+\tau_{N}) = {\rm{sgn}}\left[ \tanh \left( {I_i(t)}\right) - r_{\left[-1,+1\right]}\right] 
\label{eq:binary_stochastic_neuron}
\end{equation}
\noindent where $r_{\left[a,b\right]}$ represents a random number in the range $\left[a, b\right]$, and $\tau_{N}$ is the `neuron' time, that is the time it takes for a neuron to provide stochastic output $m_i$ with the correct statistics dictated by a new input $I_i$.

It is well-known \cite{aarts_simulated_1989} that to ensure fidelity of operation it is important to avoid \textit{simultaneous} updates of two BSNs that are  connected through a non-zero $W_{ij}$. The standard approach to avoid this issue is to update each BSN sequentially according to Eq.~\eqref{eq:binary_stochastic_neuron}, recomputing the input from Eq.~\eqref{eq:synaptic_function} after each update, a procedure known as Gibbs sampling\cite{haykin2009neural}. With sequential updating, the p-bit update rate, given as the number of flips per second ($f$), is limited by the clock speed of a given implementation. While it is possible to optimize such an approach with high-speed hardware, the sequential nature of the update limits is limiting. It is possible to enable parallel updates for groups of neurons if an encoded problem can be partitioned into decoupled groups of neurons sequenced to update independently. However, this partitioning is still constrained by the need to avoid simultaneous updates while also requiring problem specific analysis to determine the neuron update groupings. Existing hardware approaches have used problem specific partitioning to improve $f$ and we will compare our approached with these implementations in section \ref{sec:discussion}.

In contrast with sequenced BSN update approaches, the objective of this paper is to explore the feasibility of ultrafast operation through an autonomous architecture whereby each BSN continually fluctuates between -1 and +1 with probabilities that are controlled by the input $I_i$. We refer to this autonomous BSN as a \emph{p-bit} \cite{camsari_stochastic_2017} to highlight its role as the key element of an autonomous p-computer (ApC), similar to the role of a q-bit in a quantum computer. We note that such an autonomous architecture in the absence of any clocking circuitry that controls the updating p-bits has recently been demonstrated in small scale using 8 magnetic tunnel junction based p-bits\cite{borders2019integer}. With this experimental demonstration of an 8 p-bit design, it is important to understand if such a system can scale effectively.

Herein we use FPGA emulation to demonstrate the operation of a scaled version of such an autonomous computer up to 8100 (90$\times$90) p-bits with all the necessary peripheral circuitry, including programmable synapses that are used to map different problems to the co-processor. The FPGA implementation presented in this paper is specifically designed to capture the autonomous operation of probabilistic bits that fluctuate in time allowing us to make performance projections of a scaled implementation of the demonstration presented in Ref.~\cite{borders2019integer}. This paper demonstrates the feasibility of an ApC that performs the weight logic and p-bit functions defined by Eqs.~\eqref{eq:synaptic_function} and \eqref{eq:binary_stochastic_neuron} without the aid of sequencers as portrayed in Figs.~\ref{fig:apc:building_block} and \ref{fig:apc:nn_layout}. Our work is motivated by the compact, fast, energy efficient hardware that are currently being developed for the implementation of these functions \cite{hassan_low-barrier_2019} as shown in Fig.~\ref{fig:apc:mtj_apc}, which we emulate using existing CMOS devices on an easily reconfigurable, cloud accessible digital FPGA platform, Fig.~\ref{fig:apc:mtj_emulation}. 

In the following sections, we will present an emulation framework for the study of scaled autonomous probabilistic coprocessing and use the framework to provide performance predictions of a design based on nanomagnets, helping to motivate such an implementation. Section \ref{sec:apc_model} will provide an overview of the ApC, our analysis methodology, and present an abstract autonomous p-bit model with corresponding device benchmarking. Section \ref{sec:apc_emulation_framework} provides an overview of the design and implementation of a p-computing coprocessing framework using a cloud-accessible FPGA platform. Section \ref{sec:applications_model_validation} presents results for two distinct applications using the coprocessor, one involving combinatorial optimization and one involving emulated quantum annealing. Finally, sections \ref{sec:discussion} and \ref{sec:conclusion} discuss how these applications help show that accurate results can be obtained with a sequencerless probabilistic computer of the type envisioned by Feynman \cite{feynman_simulating_1982}, implemented using modern devices to enable operation at ultrafast rates unconstrained by the available clock speed in a sequenced design.

\section{Autonomous p-Computing Model}
\label{sec:apc_model}

The building block for our ApC has four components as shown in Fig.~\ref{fig:apc:building_block}:
\begin{itemize}
	\item \textit{weight logic} to implement Eq.~\eqref{eq:synaptic_function},
	\item \textit{p-bit} to implement Eq.~\eqref{eq:autonomous_equations} described below,
	\item \textit{write unit} to program the weights $W_{ij}$ and $\beta$
	\item \textit{read unit} to access the individual p-bit outputs
\end{itemize}
Fig.~\ref{fig:apc:nn_layout} shows how multiple building blocks can be interconnected to form a p-computer. The tiling shown is based on nearest-neighbor connections, but the connections need not be limited to nearest-neighbor. We have also implemented all-to-all networks using the digital emulator shown in Fig.~\ref{fig:apc:mtj_emulation}, as discussed in section \ref{sec:apc_emulation_framework}. In the following sub-sections, we will introduce what is meant by ``autonomous'' operation, present a digital model for such an autonomous p-bit, and finally benchmark the digital model against a physical device model.

\subsection{Autonomous p-bit Operation}

\begin{figure}[!t]
    \setlength\abovecaptionskip{0.1\baselineskip}
    \centering
    \includegraphics[width=\linewidth]{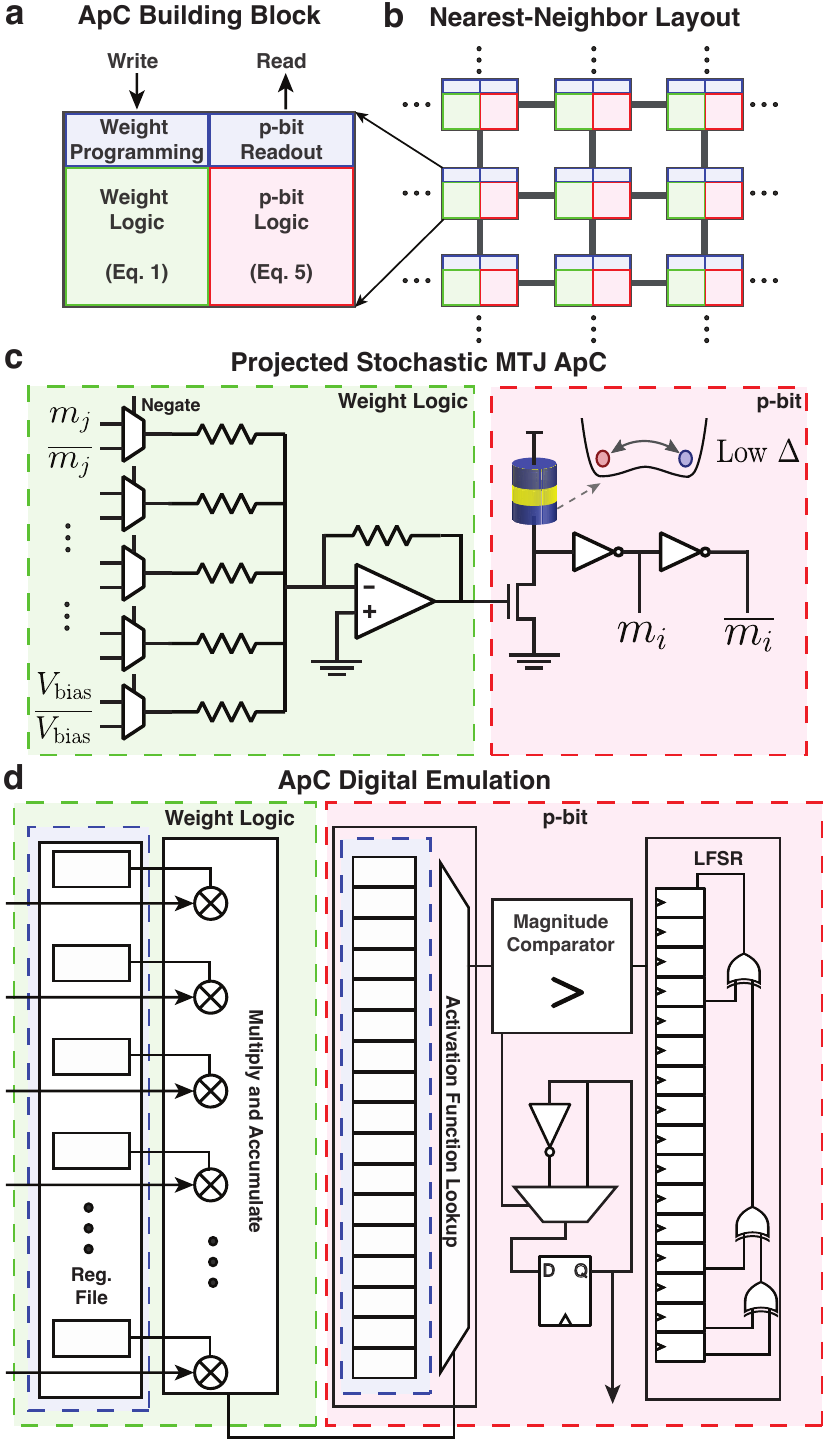} 
    \phantomsubfigure{fig:apc:building_block}
    \phantomsubfigure{fig:apc:nn_layout}
    \phantomsubfigure{fig:apc:mtj_apc}
    \phantomsubfigure{fig:apc:mtj_emulation}
    \caption{\label{fig:apc_building_block}\textbf{Autonomous p-Computer (ApC)}: {\bf a.} A weighted p-bit building block is used to construct an ApC that comprises four components supporting weight logic, p-bit logic, weight programming, and p-bit readout. {\bf b.} The individual building blocks are interconnected to construct an ApC with a desired topology. A nearest-neighbor coupling layout is depicted. {\bf c} A projected MRAM based ApC using nanomagnetic devices for the p-bit with resistor-based weight logic can be used to form a compact, efficient building block. {\bf d.} Using all-digital technology, an FPGA was used to construct and ApC that emulates the MRAM based design. An example composition of a p-bit with a linear, register based weight logic, a lookup table based activation function, a linear-feedback shift register based pseudo random number generator, and a pseudo-asynchronous attempt logic is shown.}
    \label{fig:apc}
\end{figure}

Superficially Fig.~\ref{fig:apc_building_block} looks like other existing neural network architectures, like the one used for TrueNorth \cite{merolla_million_2014}. However, to our knowledge, earlier implementations have used time-multiplexing \cite{merolla_million_2014, davies_loihi:_2018} to share the same resource among different neurons and synapses, while our objective is to eliminate sequencing and time-multiplexing altogether so that we are not constrained by available clock speeds. TrueNorth for example uses 4096 neurosynaptic cores, each core having dedicated neuron and synaptic memory forming 256 logical neurons that are time multiplexed sequentially to implement $4096 \times 256 \approx 1$M logical neurons \cite{merolla_million_2014}. 
For our sequencerless operation we would need 1M distinct building blocks for the same number of neurons. This would be impractical if we were relying on fully digital implementations, however the compact hardware implementations currently being developed makes such a design feasible. As an example, stochastic behavior of nanomagnets has recently attracted attention in the context of novel computing paradigms, and they show promise in probabilistic and neuromorphic applications \cite{fukushima2014spin,grollier_spintronic_2016, behin-aein_building_2016, liyanagedera_stochastic_2017,parks2018superparamagnetic,vodenicarevic2017low,rangarajan2017energy, lv_single_2017, mizrahi_neural-like_2018, camsari_p-bits_2019, zand_composable_2019,nasrin2019low,daniels2019energy,drobitch2019reliability,lv2019experimental}.  For example, an MRAM-based p-bit requires only 3 transistors and a magnetic tunnel junction \cite{camsari_implementing_2017}, while its digital emulation requires significantly more transistors. With such compact hardware, it is feasible to have one building block for every p-bit in order to support sequencerless operation that is not limited by clock speeds.

We call this sequencerless operation of ANNs ``autonomous''  to distinguish it from the asynchronous operation that is widely used in the context of Spiking Neural Networks \cite{merolla_million_2014,davies_loihi:_2018}.

As a quantitative measure of an ApC's speed of operation we use the number of \textit{flips} per second (\textit{f}), a \textit{flip} being defined as a p-bit update \textit{attempt} (i.e. it may choose not to actually flip). For purely \textit{sequential} updating, the number of flips per second is $\sim 1/ (\tau_{S} + \tau_{N}) $. However, as mentioned earlier, updating need not be purely sequential since unconnected BSNs can be simultaneously updated without loss of fidelity. If a number ($N_p$) out of the total number ($N$) of BSNs can be updated in parallel, then the number of flips per second will be much larger $\sim N_p / ( \tau_{S} + \tau_{N})$. Note, however, that in order to achieve this enhanced flip rate, the number of neurons that are simultaneously updated, $N_p$, have to be deliberately selected using a digital sequencer, so that the clock period, $\tau_\text{clock}$, limits the maximum number of flips per second:
\begin{subequations}
	\begin{align}
	f \leq \frac{N_p} {  \tau_\text{clock}}     \ \ \ \  \text{(Sequenced mode)}
	\label{eq:flips_per_second_sequenced}
	\end{align}
	This clock speed will be limited by the synapse and neuron times, $\tau_{S}$, $\tau_{N}$ respectively, and the overhead associated with clock distribution\ \cite{calhoun_digital_2008}.
	
	The objective of this paper is to present a framework for \emph{clockless} operation \cite{aarts_simulated_1989} whose speed is limited only by the neuron and synapse speeds
	\begin{align}
	f \leq \frac{N}{ \tau_{N}} = \frac{s N} { \tau_{S}}  \ \ \ \  \text{(Autonomous mode)}
	\label{eq:flips_per_second_autonomous}
	\end{align}
\end{subequations}
where $s \equiv \tau_{S} / \tau_{N}$. We will show that this autonomous mode provides high fidelity results without supervision keeping the fraction of detrimental simultaneous updates down to an acceptably low. Detrimental updates are managed simply by choosing a small $s$ so that  $\tau_{N}$ is much longer than $\tau_{S}$, without using a digital sequencer to enforce a deliberate update order. As a result, $f$ is not limited by $\tau_\text{clock}$ and can continue to operate faster as the synapse time $\tau_{S}$ is lowered. For example if we have nearest neighbor connections with weights of $\{-1, 0 , +1\}$, then the synapse can be implemented with short wires which respond in times less than 10 to 100 ps \cite{peng_gu_technological_2015}, much shorter than typical clock periods.

Eqs. \eqref{eq:flips_per_second_sequenced} and \eqref{eq:flips_per_second_autonomous} suggest that an autonomous design will allow faster operation (that is, more flips per second) if
\begin{align}
\tau_S  < \frac{sN}{N_p} \tau_\text{clock}
\label{eq:performance_advantage_criterion}
\end{align}
The factor $N_p/N$ represents the fraction of neurons that can be updated simultaneously, which depends on the fan-in, the nature of interconnections, and problem topology. For example, with nearest neighbor connections on a 2D square lattice as in Fig. \ref{fig:apc:nn_layout}, half the nodes can be updated simultaneously so that $N_p/N=1/2$. The factor $s$ also depends on the interconnections, but it additionally depends on the nature of the problem and the degree of solution fidelity needed, as we will show in this paper.

Ideally, we would implement a scaled version of an ApC using nanomagnetic devices to explore the performance of such a design. However, given current technological limitations, we will model the operation of p-bit hardware using a digital FPGA platform. In the following section we discuss how we use a digital, synchronous device to emulate intrinsically asynchronous nanomagnets.

\subsection{Autonomous p-bit Model}
As digital platforms are inherently synchronous, we mimic autonomous operation by replacing Eq.~\eqref{eq:binary_stochastic_neuron} with a new hardware-inspired model, Eq.~\eqref{eq:autonomous_equations} below that we benchmarked against established device models (section \ref{sec:benchmarking_llg}). These equations are based on SPICE simulations of Boltzmann networks where the update order of p-bits becomes irrelevant due to the symmetric coupling between connected p-bits. Such a clockless circuit corresponds to the \textit{asynchronous parallelism} scheme {used} to realize Boltzmann Machines in hardware with no asymptotic guarantees for convergence \cite{aarts_simulated_1989} unless all p-bits operate with up-to-date information that is enabled by fast synapses \cite{pervaiz_hardware_2017}. {This model is valid for such networks, however, for other networks such as those with directed connections, the update ordering of p-bits may be important and other hardware models more appropriate for these systems are likely required and are not discussed in this paper.}

At each time step, all p-bits are free to flip and they do so with a probability $\sim s$ that is controlled by the input $I_i$ having a zero-input value $s(I_i = 0) = s_0 \ll 1$.
\begin{subequations}
	\begin{align}
	m_i (n+1) = m_i(n) \times {\rm{sgn}}[ e^{-s}  - r_{\left[0,1\right]}  ]
	\label{eq:autonomous_equations:a}
	\end{align}
	\vspace{-0.2in}
	\begin{align}
	s = s_0 e^{-m_i(n) I_i(n)}
	\label{eq:autonomous_equations:b}
	\end{align}
	\label{eq:autonomous_equations}
\end{subequations}
{As each p-bit flips with a probability $\sim s$ in each time ste}p, the average time taken for a p-bit to respond is $1/s$. Since time steps are measured in units of $\tau_{S}$, we have $\tau_{N} = (1/s) \times \tau_{S}$ as stated earlier. Unlike Eq.~\eqref{eq:binary_stochastic_neuron}, Eq.~\eqref{eq:autonomous_equations} can be used to update all p-bits in parallel without explicitly worrying about simultaneous updates. With small values of $s_0$, the fraction of simultaneous updates is sufficiently small such that Eq.~\eqref{eq:autonomous_equations} in an unsequenced mode gives results equivalent to those obtained from careful sequencing using Eq.~\eqref{eq:binary_stochastic_neuron}. 

For a given network topology and embedded problem\cite{choi_minor-embedding_2008, choi_minor-embedding_2011}, the value of $s$ that ensures convergence to thermal equilibrium must be identified in order to assess the amount of parallelism in the design. The desire is to find the smallest value of $s$ that converges the thermal equilibrium for the problem specific Boltzmann distribution. {In section \ref{sec:applications_model_validation} we explore} two example problems and evaluate the degree of parallelism obtainable. For example, in the nearest-neighbor design of Fig. \ref{fig:max_cut}, a value of $s = 1/4$ ensures convergence for a network of 8100 neurons. This value of $s$ states that on average the number of neurons that update within each synapse delay is $N \times s = 2025$. However, as the problem complexity increases, i.e. the incorporation of on-site biases, the acceptable value of $s$ decreased to $1/12$ as shown in Fig. \ref{fig:quantum_annealing}. Ultimately, the term $s$ drives the physical design of the synapse and neuron implementation.

\subsection{Model Benchmarking with Stochastic LLG}
\label{sec:benchmarking_llg}
{The autonomous p-bit model of \eqref{eq:autonomous_equations:a} and \eqref{eq:autonomous_equations:b} was benchmarked against a coupled stochastic Landau-Lifshitz-Gilbert (sLLG) equation}. Magnetization dynamics of the modeled circular stochastic nanomagnet are captured by solving the sLLG equation \cite{butler_switching_2012}:
\begin{subequations}
	\begin{align}
	&(1+\alpha^2)\frac{d\hat m}{dt} = -|\gamma|{\hat m \times \vec{H}} - \alpha |\gamma| (\hat m \times \hat m \times \vec{H})\nonumber \\ &+  \frac{1}{q  N_s}(\hat m \times \vec{I}_{S} \times \hat m)  + \left(\frac{\alpha}{q N_s} (\hat m \times \vec{I}_{S})\right)
	\label{sLLG}
	\end{align}
\end{subequations}
\noindent where $\alpha$ is the damping coefficient, $\gamma$ is the electron gyromagnetic ratio, $N_s=\rm M_s$Vol./$\mu_B$ is the total number of Bohr magnetons in the magnet, $\rm M_s$ is the saturation magnetization, $\vec{H}=\vec{H_d}+\vec{H_n}$ is the effective field including the out-of-plane ($\hat x$ directed) demagnetization field $\vec{H_d}=-4\pi M_s m_x \hat x$, as well as the thermally fluctuating magnetic field due to the three dimensional uncorrelated thermal noise $H_n$  with zero mean $\langle{H_n}\rangle=0$ and standard deviation $\langle H_n^2\rangle= 2\alpha \rm kT / |\gamma| M_sVol.$ along each direction, $\vec{I_S}$ is the applied spin current to the nanomagnet. The HSPICE solver we employed is based on spherical coordinates that solves for ($\theta$,$\phi$), but the noise is first included as three uncorrelated random magnetic fields in Cartesian coordinates before being turned into spherical coordinates \cite{torunbalci2018modular}. The HSPICE solver uses the \verb|.trannoise| function\cite{torunbalci2018modular} and is benchmarked against our own MATLAB implementation that uses the Stratonovich convention \cite{behin2010proposal}, as well as the Fokker-Planck Equation \cite{butler_switching_2012, torunbalci2018modular}. We have used a time step of ~ 1 ps for the chosen parameters which is verified by comparing the equilibrium fluctuations of single nanomagnets that are obtained numerically, against the expected Boltzmann distribution. \\

Normally, a nanomagnet based p-bit thresholds its continuous output {with} an inverter \cite{camsari_stochastic_2017,camsari2017implementing} that is typically included in SPICE simulations with additional transistors. In order to simplify numerical simulations in the present context, we artificially threshold
the sLLG outputs (such that $m_z>0 \rightarrow 1$ and $m_z<0 \rightarrow 0$) at each time step. This allows a binarization of the sLLG outputs that allow each p-bit to have a binary output, according to Eq.~\ref{eq:binary_stochastic_neuron}. Note that in an actual device implementation, a single inverter is enough to threshold the output given that a moderate tunneling magnetoresistance value (TMR=$R_{AP}/R_P-1 \approx 100\%$) leads to voltage fluctuations of $\approx  200$ mV's over a voltage division of $V_{DD}=0.8$ V, which is more than enough for a single inverter to threshold these fluctuations to rail-to-rail voltages \cite{camsari2017implementing}. For a detailed analysis of the nanomagnet based p-bit design that includes device-to-device variations, see Ref~\cite{borders2019integer}.

Individual p-bits are coupled according to:

\begin{equation}
I_{s,i}^z (t+\Delta_t) = \beta I_{s0} \ \sum_{j} W_{ij} \text{ sgn}(m_j^z(t) )
\label{eq:sLLG_coupling}
\end{equation}
where, $I_{s0}$ is the $\tanh$ fitting parameter of the sigmoidal response ($\text{sgn}(m_z)$ versus spin current $I_s^z$ along $z$-direction). Equation \ref{eq:sLLG_coupling} and Equation ~\ref{eq:autonomous_equations} constitute the autonomous behavioral model that is used to benchmark the results of coupled sLLG equations, which we refer to as ``behavioral model'' hereon. In the benchmark, a circular disk magnet with a vanishing anisotropy ($H_K$) is used  with the parameters: diameter $D=150$ nm and thickness $t=2$ nm, $\alpha=0.01$, $M_s=1100$ emu/cc, $H_K=1$ Oe resulting in an autocorrelation time of $\tau_{corr}=1.372$ ns and $I_{s0}=1$ mA. A fitting parameter of $1.4$ is used in the behavioral model for $\tau_{N}$, i.e. $\tau_{N}=1.4\tau_{corr}$. 

\begin{figure}[!tb]
	\centering
	\includegraphics[width=\linewidth]{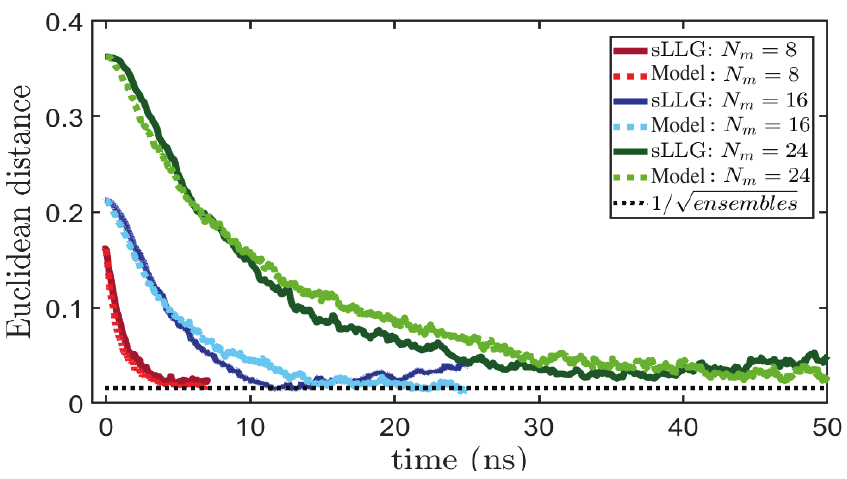} 
	\caption{\textbf{Benchmarking the behavioral model with sLLG using Euclidean distance}: Using a random Sherrington-Kirkpatrick spin glass instance for different network sizes, $N$, the behavioral model is benchmarked against sLLG as a function of time. Each point on the graph represents the Euclidean distance from the ideal Boltzmann distribution and the ensemble solution obtained from the behavioral model and sLLG. The steady state error will depend on the number of ensembles as shown by the black dotted line.}
	\label{fig:ED}
\end{figure}

{We use a simulated Sherrington-Kirkpatrick \cite{sherrington_solvable_1975} spin glass network with a random coupling matrix and random bias between -1 and +1.} The benchmarking of the proposed behavioral model with the coupled sLLG network, analogous to the probabilistic circuit  proposed in \cite{camsari2017stochastic}, is accomplished by comparing two different quantities: (1) Euclidean distance and (2) Free energy.

Euclidean distance is defined by:
\begin{equation}
ED=\sqrt{\sum_{i=1}^{2^{N}}(P_i-P_{\text{i,Boltzmann}})^2}
\label{eq:ED}
\end{equation}
where $P_i$ is the probability of occurrence of the $i$-th configuration computed out of 4000 ensembles at each time step of the simulation. $P_\text{i,Boltzmann}$ is computed from the joint probability distribution obtained from a Boltzmann law. 

\begin{figure}[!tb]
	\centering
	\includegraphics[width=\linewidth]{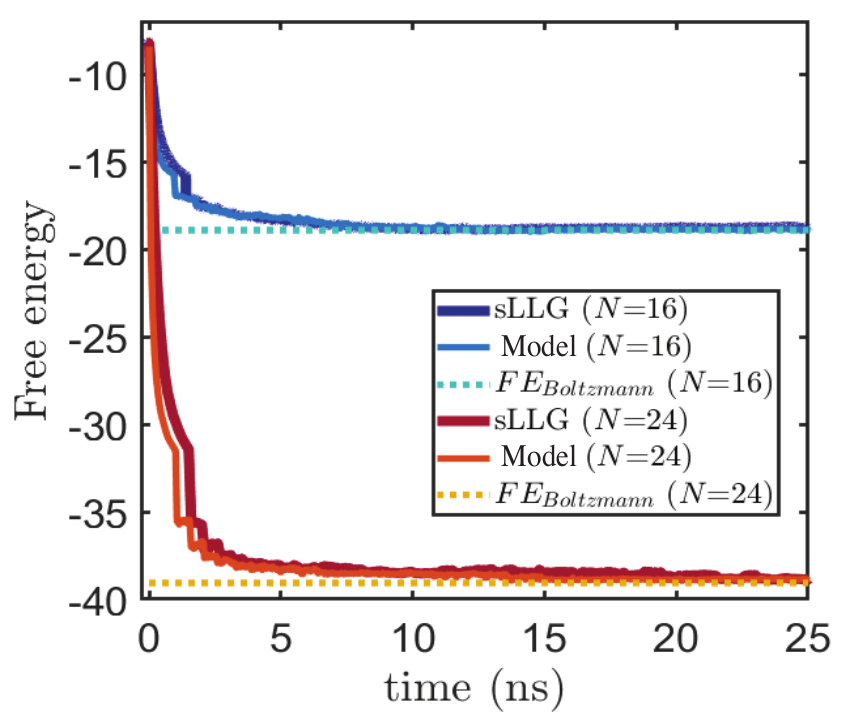} 
	\caption{\textbf{Benchmarking the behavioral model with sLLG using Free Energy}: The free energy calculated for the random Sherrington-Kirkpatrick spin glass instance of Fig.~\ref{fig:ED} from the behavioral model is benchmarked against sLLG as a function of time for network sizes $N=16$ and $N=24$, showing convergence to the free energy obtained from Boltzmann law.}
	\label{fig:free_energy}
\end{figure}

The second benchmark approach is based on a comparison of the free energy of the system with what is expected from the principles of statistical mechanics. Free energy is defined by \cite{reichl1999modern} the partition function $Z$:
\begin{equation}
FE=\frac{\ln(Z)}{-\beta}`
\label{eq:FE}
\end{equation}
where, $\beta$ is the pseudo-inverse temperature. Partition function $Z$ is given by:
\begin{equation}
Z=\sum_k{\exp(-\beta E_k)}
\end{equation}
where $k$ represents different configurations of the network. Energy of a specific configuration is defined by:
\begin{equation}
E_k=-0.5\sum_{\substack{i,j \\ i\neq j}}W_{ij}m_im_j-h_im_i-h_jm_j
\end{equation}

When numerically calculating free energy from the sLLG data, the following steps have been applied (similar to the importance sampling method described in \cite{liu2015estimating}):
\begin{enumerate}
	\item The probability of different configurations, $P_i$, are calculated out of 4000 ensembles for each time step
	\item For each $P_i$ larger than a certain threshold value $P_{th}$, the partition function $Z_i=\exp(-I_0E_i)/P_i$ is calculated, so that outliers are excluded
	\item For each $Z_i$, the free energy $FE_i=-\ln(Z_i)/I_0$ is calculated.
	\item Finally the mean of all $FE_i$ is computed.
\end{enumerate}

The above method is suitable for small examples, {but may not scale due to the difficulty in empirically calculating different probabilities $P_i$ as the network size grows. The striking agreement between the sLLG model and the behavioral model given by Eq.~\eqref{eq:autonomous_equations} shown in Fig.~\ref{fig:ED} and Fig.~\ref{fig:free_energy} helps establish the validity of} Eq.~\ref{eq:autonomous_equations} as a suitable digital model of an autonomous, stochastic MTJ-based computer.

\section{Emulation Framework}

{Having established a model for the autonomous p-bit operation, in this section we describe the design and implementation of an FPGA based framework to explore the performance, scalability, and other characteristics of an ApC.}

\label{sec:apc_emulation_framework}
\subsection{Autonomous FPGA Coprocessor}
\label{sec:autonomous_fpga_emulator}
\begin{figure}[!tb]
	\setlength\abovecaptionskip{-0.5\baselineskip}
	\centering
	\includegraphics[width=\linewidth]{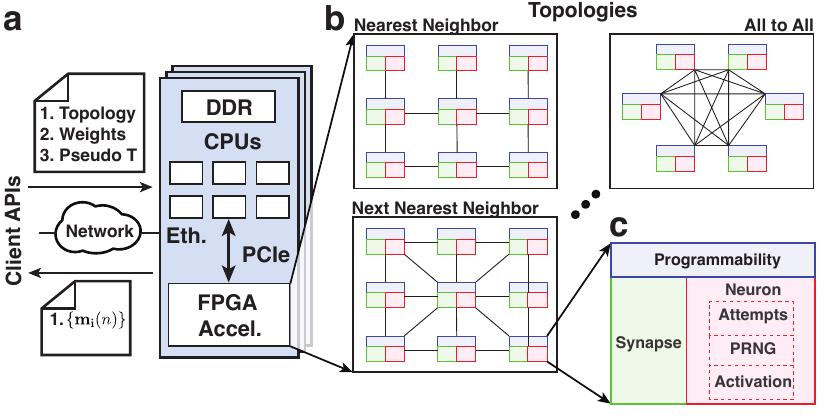} 
	\phantomsubfigure{fig:sup_pcomputing_architecture:a}
	\phantomsubfigure{fig:sup_pcomputing_architecture:b}
	\phantomsubfigure{fig:sup_pcomputing_architecture:c}
	\caption{\label{fig:sup_pcomputing_architecture} {\bf Cloud Accessible $p$-computing Co-processor} {\bf a.} Client applications are used to specify a desired network topology, problem specific weights, and current pseudo-temperature through a network accessible pool of dedicated servers. The servers provide general purpose processing, network connectivity, and PCIe accessible FPGA accelerators. After loading the desired topology into the FPGA, the system operates and provides a sampling interface for the current state of the network p-bits. {\bf b.} The architecture supports multiple topologies depending on the problem of interest ranging from nearest-neighbor connectivity to all-to-all connectivity. {\bf c.} Each of the modular p-bits in the network comprises a bus accessible programmable interface, synapse block, and neuron. The neuron is further modularized to support different approaches for flip-attempt logic, pseudo random number generation, and activation function implementations.}
\end{figure}

An all-digital framework based on Eqs.~\eqref{eq:autonomous_equations:a} and \eqref{eq:autonomous_equations:b} was developed, Fig.~\ref{fig:sup_pcomputing_architecture},  to facilitate architectural exploration of an ApC, study various trade-offs in p-bit, weight logic, and topology design, and to accelerate the combinatorial optimization and sampling problems explored in section \ref{sec:applications_model_validation}. The digital framework leverages reconfigurable computing devices to support rapid exploration of different designs. A Xilinx Virtex Ultrascale$+$ \verb|xcvu9p-flgb2104-2-i| provided via Amazon Web Services F1 cloud-accessible EC2 compute instances was used for the ApC implementation. While a Xilinx FPGA was used in this work, the design is hardware agnostic and another device, e.g. an Intel Stratix FPGA, {could} readily be used.

As shown in Fig.~\ref{fig:apc:nn_layout} and Fig.~\ref{fig:sup_pcomputing_architecture:b}, an ApC comprises multiple weighted p-bits arranged in various topologies, each supporting programmable problem instances. There are many options for the implementation of programmable control, weight logic, p-bits, and p-bit readout in a digital platform. The digital ApC of Fig.~\ref{fig:sup_pcomputing_architecture} comprises a modular weighted p-bit, Fig.~\ref{fig:sup_pcomputing_architecture:c}, that can be organized into various topologies, Fig.~\ref{fig:sup_pcomputing_architecture:b}, supporting programmed problem instances. An example weighted p-bit implementation is shown in Fig.~\ref{fig:apc:mtj_emulation} leveraging a memory-mapped weight-logic register bank supporting linear weight coupling. The output of the weight logic block is provided to a programmable, activation function look-up table that is used in conjunction with a Linear Feedback Shift Register (LFSR) based pseudo-random function to implement Eqs.~\eqref{eq:autonomous_equations:a} and \eqref{eq:autonomous_equations:b}. Additional options were developed and explored for these building block elements, see section \ref{sec:digital_building_blocks} and Fig.~\ref{fig:sup_digital_options}, beyond what is shown in Fig.~\ref{fig:apc:mtj_emulation}.

Interaction with the FPGA framework is provided through MATLAB MEX programs in a client-server command driven model, Fig.~\ref{fig:sup_pcomputing_architecture:a}. Clients issue commands to select which of the pre-built topologies to program into the cloud FPGA instance, the current pseudo-temperature for the network, problem specific weights, and options to pause or resume the network. Commands are also provided to support random sampling from the network for readout operation. Online annealing is directly supported through global update operations of the activation function look-up. All weights are dynamically programmable through memory-mapped operations. The server interfaces with a {PCI Express (PCIe)} attached FPGA, interacting with the programmed design as commanded. Other clients such as Octave and Python are readily supported through a \lstinline{C++} abstraction layer leveraging networking and serialization libraries.

\begin{figure*}[t!]
	\setlength\abovecaptionskip{\baselineskip}
	\centering
	\includegraphics[width=\linewidth]{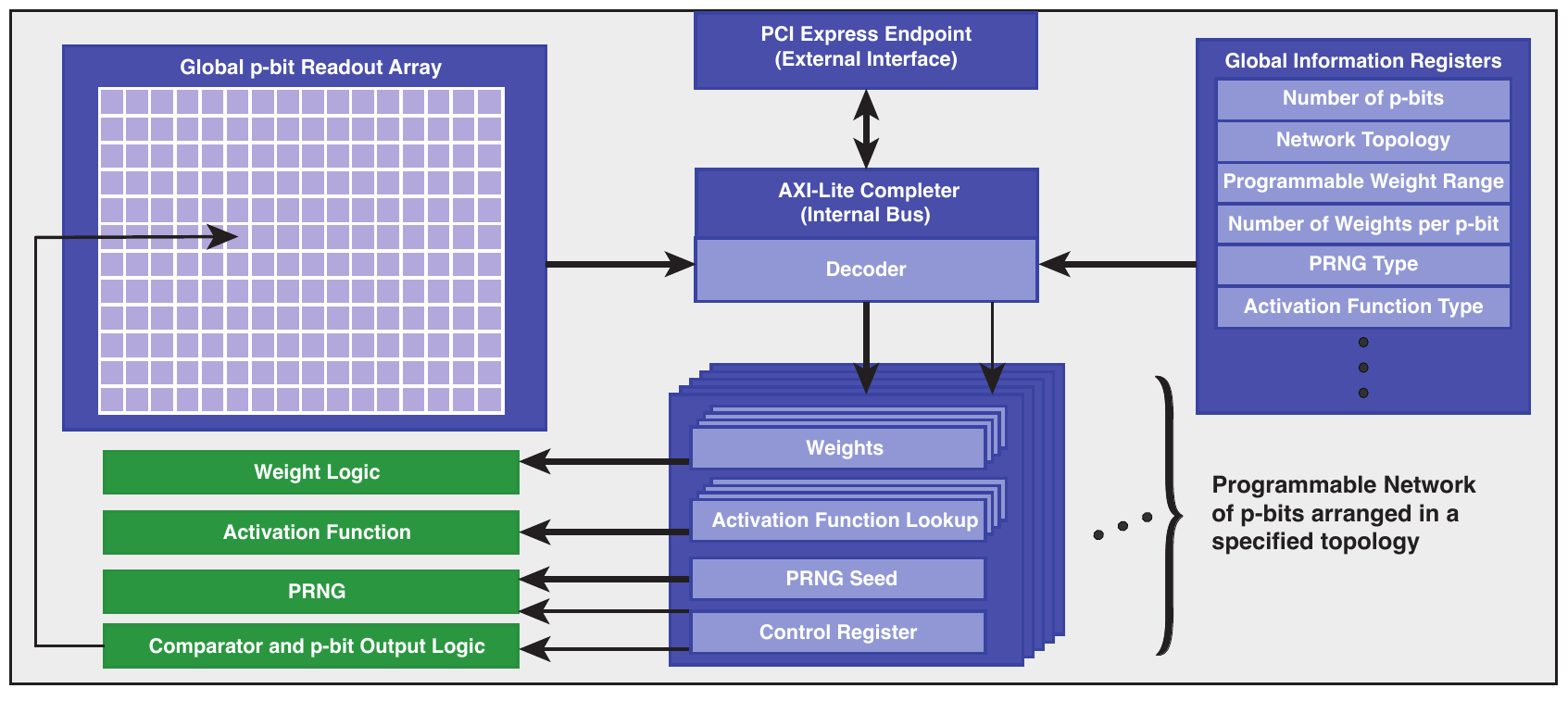} 
	\caption{\label{fig:sup_mmap} \textbf{Logical Organization and Memory Map}: The FPGA based ApC coprocessor is accessible to a host processor from a dedicated PCIe endpoint within the FPGA. This endpoint provides logical access to the internal memory map of the ApC. The ApC contains global information and control registers, a global p-bit array facilitating readout, and finally individual control of each weighted p-bit through a localized, dedicated address space.}
\end{figure*}

Fig.~\ref{fig:sup_mmap} depicts the logical organization of the FPGA ApC design used herein. All interaction with the FPGA is performed using a PCIe Gen3 x16 interface supporting direct memory access (DMA) transactions {(requester)}. Programmable control of the design is accomplished using an {AXI-Lite 32-bit completer interface} and memory map decoder. The address space for the ApC is divided into a few regions: global control and information registers, p-bit readout array, and individual control for each weighted p-bit in the design.

The global address space provides information on the ApC pertinent for client interaction with the system. This includes information such as the total number of p-bits in the design, the p-bit network topology, weight precision, and other useful run-time information such as the number of elapsed synapse delays between client sampling requests. Additionally, global control functions include the ability to pause and resume the network so that a client can access the global p-bit readout array. This readout array holds the current output of all p-bits sampled on each system clock cycle. Reads from this interface are performed when the network is paused to ensure atomic readout of all p-bits given the limited bandwidth of the readout interface.

Each p-bit in the system has a local memory space supporting programmable control of its function. Each p-bit has localized programmable weights enabled through registers or internal memory (RAM), a programmable activation function lookup table supporting direct look-up or interpolation, support for seeding the chosen pseudo random number generation (PRNG) function, and finally a set of control registers for the p-bit. The output of the p-bit programmable elements directly interface with the weight logic, activation function, PRNG, and comparator operation of the weighted p-bits. Online annealing is accomplished using a global bus broadcast when programming the activation function lookup tables, so that all p-bits are updated simultaneously. Alternatively, each p-bit's activation lookup table can be independently controlled, allowing the exploration of non-uniform bias, local temperature effects, and other non-idealities, facilitating future opportunities for exploration.

\subsection{Digital Building Blocks}
\label{sec:digital_building_blocks}

\begin{figure*}[t!]
	\setlength\abovecaptionskip{\baselineskip}
	\centering
	\includegraphics[width=\linewidth]{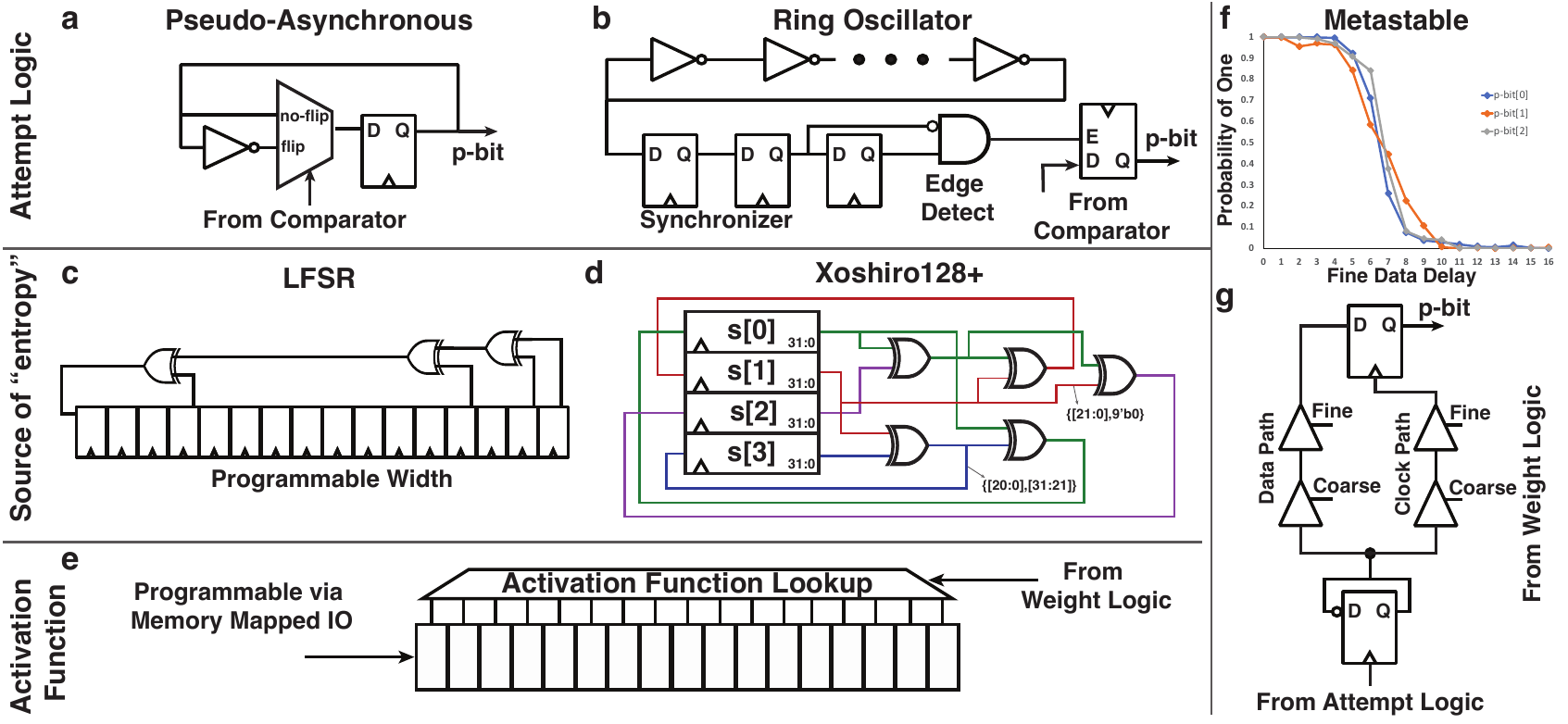} 
	\phantomsubfigure{fig:sup_digital_options:a}
	\phantomsubfigure{fig:sup_digital_options:b}
	\phantomsubfigure{fig:sup_digital_options:c}
	\phantomsubfigure{fig:sup_digital_options:d}
	\phantomsubfigure{fig:sup_digital_options:e}
	\phantomsubfigure{fig:sup_digital_options:f}
	\phantomsubfigure{fig:sup_digital_options:g}
	\caption{\label{fig:sup_digital_options} {\bf Modular Autonomous Digital p-bit Components:} {\bf a.} Using output from a comparator, the state of a given p-bit is probabilistically flipped on any given system clock edge. {\bf b.} Alternatively, a free-running ring oscillator can be used to generated edges centered around a oscillator characteristic frequency, mimicking the attempt rate of a stochastic nanomagnet. These edges control the enable of the p-bit which is then updated based on the output of a comparator. {\bf c.} Pseudo-randomness is used to provide the \lstinline{rand()} function. A linear feedback shift register is an area efficient means to generate pseudo random values, however the quality is limited. {\bf d.} For improved quality of pseudo random output at the expense of more area, a Xoshiro128+ \cite{blackman_scrambled_2018} generator can be used. {\bf e} The p-bit activation function can be implemented using a straight-forward look-up table or through an interpolated look-up table output. {\bf g.} By combining the source of ``entropy'' with the activation function, a tunable delay chain can be used to leverage controlled metastability at the input of a flip-flop to produce a probabilistic output. Outputs from three independent p-bits are shown overlaid in {\bf f}.}
\end{figure*}

A modular digital p-bit was designed to support different options for the p-bit building block elements. Each p-bit is logically partitioned into a unit for pseudo-randomness or ``entropy'', a block for computing the activation function, and a portion that uses the results of a comparison between the activation function and PRNG to determine if a p-bit update attempt should occur. There are various ways to construct these elements using digital logic. In this design, a few select implementations were explored as shown in Fig.~\ref{fig:sup_digital_options}. {A non-exhaustive exploration of design options described below was performed as part of initial trade-study. As quantitative results were not pursued, a qualitative assessment of the various options is provided for completeness.}

Fig~\ref{fig:sup_digital_options:a} and \ref{fig:sup_digital_options:b} show two methods for performing autonomous updates of each p-bit. Shown in Fig. \ref{fig:sup_digital_options:a} is the logic corresponding to Eq.~\eqref{eq:autonomous_equations:a} where the comparator provides $\text{sgn}\left(e^{-s} - r_{\left[0,1\right]}\right)$. This approach emulates autonomy while preserving fully synchronous operation of the digital design. This p-bit update logic was used for the problems in this work. The activation look-up is programmed with $e^{-s}$.

A single clock domain within the FPGA is used to synchronize the digital elements of each synapse and p-bit. For each global clock period, $t_{ck}$, the synapse logic requires $\tau_{s} = N_{syn} t_{ck}$ time to compute where $N_{syn}$ is the number of global clock events required. In the nearest-neighbor models, it is possible for $N_{syn}$ to need only a single cycle, however, as the the weight precision and number of neighbors is increased, $N_{syn}$ also increases. Thus, when specifying an $s$ ratio, coupled with a design driven $N_{syn}$, the neuron time, encoded probabilistically in the look-up table, is $\tau_{N} = s^{-1} N_{syn} t_{ck}$.

Alternative approaches were explored and implemented for the p-bit updates including the use of free-running ring oscillators, Fig.~\ref{fig:sup_digital_options:b}, to mimic the naturally stochastic update frequency of an MTJ based p-bit as described by Eq. \eqref{eq:binary_stochastic_neuron}. In this implementation, each p-bit has a dedicated free running ring oscillator that generates asynchronous ``attempt'' edges that are synchronized into a system clock domain. Each asynchronous edge determines when the p-bit should attempt to update based on the current output from a comparator according to Eq.~\eqref{eq:binary_stochastic_neuron}, in which case the activation look-up is programmed with $\tanh$. {While the use of oscillators provides some degree of true randomness for when flip attempts occur, this benefit was marginal and did not out weigh the design complexity and overhead (e.g. area and power) introduced with their use. As a result, the design of Fig. \ref{fig:sup_digital_options:a} was selected for the attempt logic.}

While the attempt logic is used to determine \emph{when} an update attempt should be made, the logic relies on the output of a comparator to determine \emph{if} a flip should occur. The comparator computes the sign of the difference between the output of a PRNG and the output of the programmed activation function. Shown in the second row of Fig.~\ref{fig:sup_digital_options} are two PRNG implementations. A Linear Feedback Shift Register (LFSR) is a PRNG that provides pseudo randomness in a compact design at the expense of output quality, Fig. \ref{fig:sup_digital_options:c}. While the LFSR may be sufficient for many problems, a higher-quality PRNG was implemented that requires minimal FPGA resources, but significantly improves the PRNG quality \cite{blackman_scrambled_2018}. 

The second input to the comparator is from an activation function output, shown on the third row if Fig.~\ref{fig:sup_digital_options:e}. As implemented, a straight-forward look-up table was sufficient for the designs in this paper. However, an improved interpolation based activation function\cite{ortega-zamorano_high_2014} would improve the accuracy of the lookup results and may be necessary for certain problem classes.

An additional building block was created to leverage \textit{physical} randomness and a built-in sigmoidal response from within a digital design as shown in Fig.~\ref{fig:sup_digital_options:g}. Leveraging a delay based building block \cite{majzoobi_fpga_2010} and flip-flop metastability, ``true'' entropy was used to construct a sigmoidal response as depicted in Fig.~\ref{fig:sup_digital_options:f}. However, over continued operation within the device, temperature and other variations caused the sigmoidal curves to drift, resulting in non-uniform bias and operation. As a result, this building block is not currently being used in the design; however, it does provide insight into non-idealities that may be encountered in a chip design.

Finally, we note that an ApC that emulates the physics of different hardware primitives including memristive \cite{mahmoodi2019versatile} stochastic neurons could be implemented in an autonomous circuit, unlike the nanomagnet inspired equations Eq.~\ref{eq:autonomous_equations:a}-Eq.~\ref{eq:autonomous_equations:b}. {Additionally, as discussed earlier, we do not explore directed networks in this work, however the overall FPGA architecture was designed to support the exploration of different hardware models and network topologies, hence they can be included here with minimal effort in the future.}

\section{Applications and Validation}
\label{sec:applications_model_validation}

{In this section we explore two applications of the ApC using the FPGA framework: combinatorial optimization and quantum emulation.}

\subsection{Combinatorial Optimization}

\begin{figure}[!t]
    \setlength\abovecaptionskip{-0.5\baselineskip}
    \centering
    \includegraphics[width=1\linewidth]{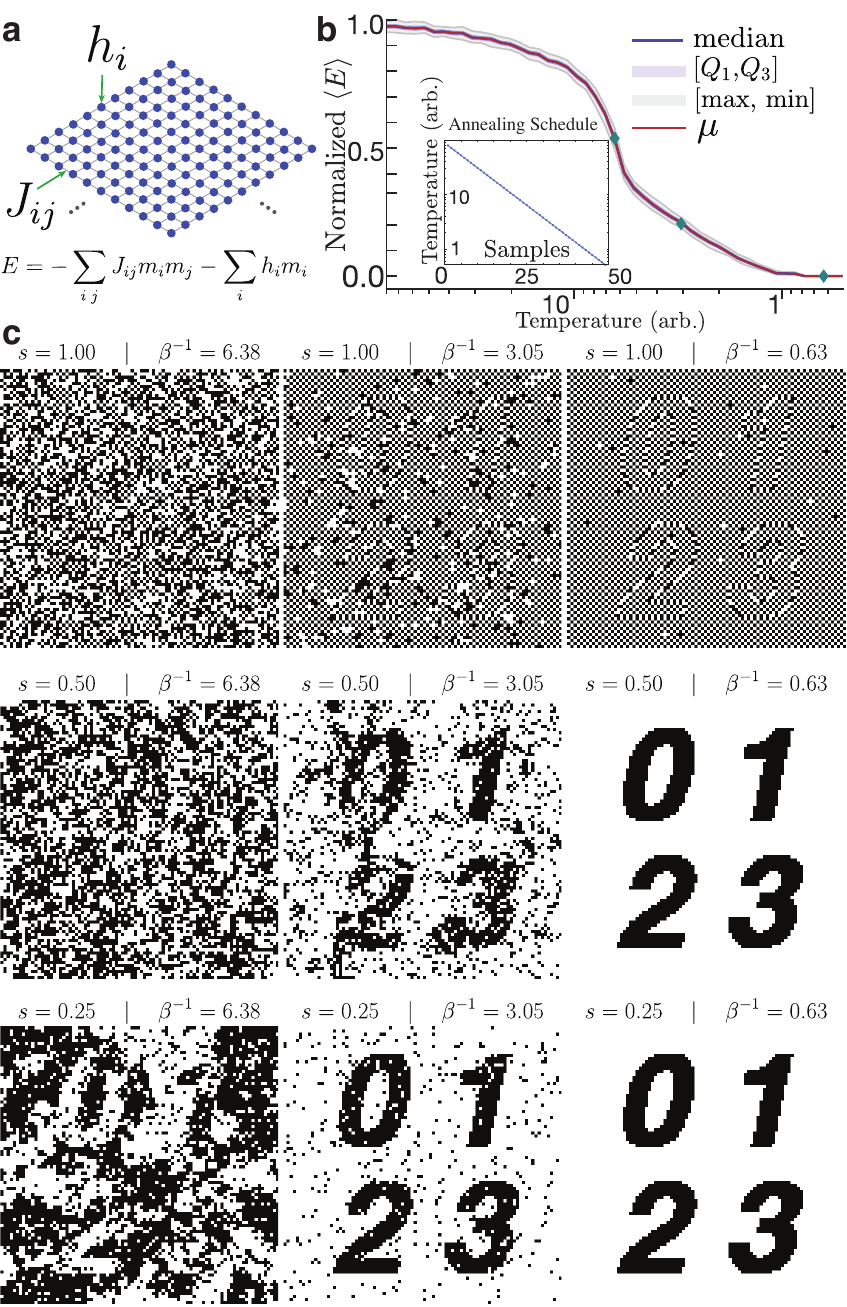} 
    \phantomsubfigure{fig:max_cut:a}
    \phantomsubfigure{fig:max_cut:b}
    \phantomsubfigure{fig:max_cut:c}
    \phantomsubfigure{fig:max_cut:d}
    \phantomsubfigure{fig:max_cut:e}
    \caption{\label{fig:max_cut}{\bf Combinatorial Optimization: Max-Cut problem} {\bf a.} An ~8K spin ($90 \times 90$) nearest-neighbor Ising lattice of p-bits supporting $\{\pm 1, 0\}$ weights can be used to solve a Max-Cut combinatorial optimization problem. Here a black and white image is used to generate magnetic domains corresponding to each character. The problem is then solved leveraging an FPGA ApC co-processor (see Methods). {\bf b.} Online annealing is used to transition the network from a high to low temperature through a reduction of $T(n+1) = 0.9\ T(n)$ where $T = \beta^{-1}$ (inset). During this process the network converges to a low energy state solution to the problem of interest. {\bf c.} The network features begin to emerge as lower temperatures are reached. The probability of an individual p-bit flipping is controlled to explore how simultaneous updates effect the network or $s=1$ {\bf c}, $s=1/2$ {\bf d}, and $s=1/4$ {\bf e}. If $s$ is too large, there is no convergence to the ideal solution. As $s$ is lowered, the network is more effective at finding low energy solutions, even at higher temperatures.}
\end{figure}


A common architecture used to solve combinatorial optimization problems is the nearest-neighbor Ising model as shown in Fig. \ref{fig:max_cut:a}. In the Ising model, each p-bit is connected to its neighbor through a coupling matrix, $J_{ij}$, and is influenced by an on-site bias $h_i$. Note that this is trivially mapped into $W_{ij}$ as in Eq.~\eqref{eq:synaptic_function}. By mapping a problem of interest to this system, an Ising computer will intrinsically search for the lowest energy solution to the problem.

Typical implementations of these systems leverage some form of simulated annealing in hardware to guide the system into a low energy solution, using careful control and sequencing of spin-flip updates to avoid non-ideal spin updates \cite{baity-jesi_janus_2014, yamaoka_20k-spin_2016, yoshimura_fpga-based_2016}. In the context of a nearest-neighbor topology, the network can be split into two groups in a checkerboard-like pattern such that all spins in a given group can be updated in parallel \cite{ortega-zamorano_fpga_2016}. This results in the ability to update half of all spins within a given clock period. According to Eq.~\eqref{eq:flips_per_second_sequenced} this results in 
\begin{equation}
f = \frac{N}{2\tau_{clock}}
\label{eq:ising_nn_checkerboard}
\end{equation} 

An ApC can be used to solve the same class of problems, but to do so a value of $s$ must be found that produces a solution with the desired fidelity. By identifying this limit for $s$, an upper bound is placed on the synapse speed that must be obtained to achieve a competitive $f$ from Eq.~\eqref{eq:performance_advantage_criterion}.

Shown in Fig.~\ref{fig:max_cut} is a Max-Cut problem for which a black and white image was used to encode magnetic domains for the p-bits. The network is initialized to run at a high effective-temperature (low $\beta$) resulting in an effectively uncoupled network. The temperature is then gradually reduced according to an annealing scheduling until the network crystallizes at a low energy, in this case ideal, solution as shown in Fig.~\ref{fig:max_cut:b}. Shown in Figs. \ref{fig:max_cut:c}, \ref{fig:max_cut:d}, \ref{fig:max_cut:e}, different values of $s$ are used to convey how simultaneous updating affects the ability of the network to converge to the solution. As $s$ is decreased from $1$ to $1/2$ and $1/4$, the smaller values of $s$ result in a more effective convergence. While these results are heuristic in nature given their visual display, a quantitative energy based analysis conveys the same relationship as discussed in the following section with a demonstration of simulated quantum annealing.

For this problem, a value of $s = 1/4$ resulted in effective convergence to the ideal solution and well-behaved network operation. Given this result, the ApC has
\begin{equation}
f = \frac{N}{4\tau_{S}}
\label{eq:ising_nn_apc}
\end{equation} 
Comparing \eqref{eq:ising_nn_checkerboard} and \eqref{eq:ising_nn_apc}, as long as the synapse is twice as fast as the best $\tau_\text{clock}$ that could be obtained from a synchronous implementation, the network will perform the same number of flips per second without needing a sequencer.

\subsection{Quantum Emulation}

Simulating the behavior of quantum systems using classical models has long attracted interest.  After its introduction by Ref.~\cite{suzuki1976relationship}, it has been recognized that thermodynamic features of quantum systems that avoid the ``sign problem'' \cite{troyer2005computational}
can be efficiently simulated by classical computers. This allows Quantum Annealing (QA) algorithms designed for quantum systems to be simulated on classical computers, an approach that is called Simulated Quantum
Annealing (SQA). Computationally, SQA uses a finite number of ``replicas'' where each replica is sized to match the size of the original quantum system. This replication enables a mapping of the quantum system to a classical collection of p-bits.  The number of replicas that are needed depends on the temperature of the quantum system \cite{santoro2002theory,heim2015quantum} and the desired accuracy to emulate the quantum system. SQA algorithms are typically run on software \cite{santoro2002theory,heim2015quantum,denchev2016computational} and on sequencer-based hardware designs \cite{hitachi_sqa_2017,waidyasooriya_opencl-based_2019}. Here, we demonstrate how quantum systems can be emulated using our ApC by solving a model quantum system, the Transverse Ising Hamiltonian, and establish how ApC exactly reproduces the thermodynamics of a many-body quantum system at finite temperatures and magnetic fields.  

\begin{figure}[!t]
  \centering
 \setlength\abovecaptionskip{-0.5\baselineskip}
 \includegraphics[width=1\linewidth]{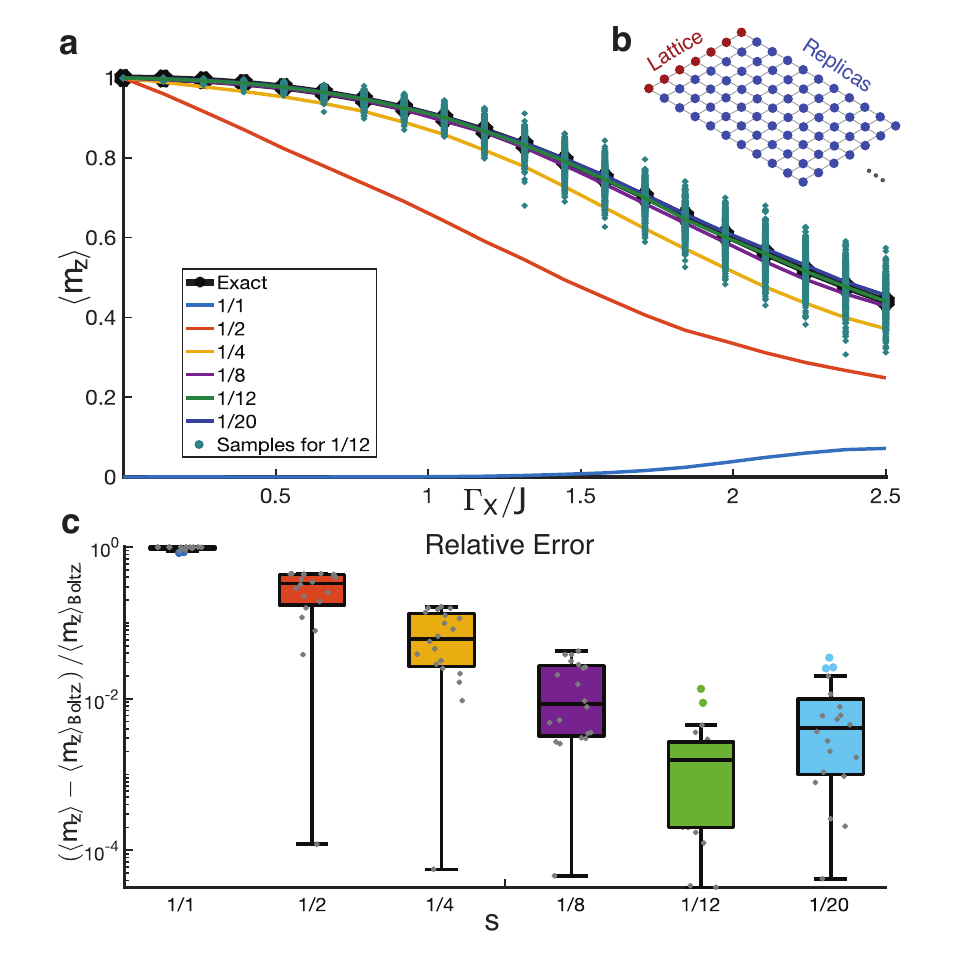}
 \phantomsubfigure{fig:QA:a}
 \phantomsubfigure{fig:QA:b}
 \phantomsubfigure{fig:QA:c}
 \caption{\label{fig:quantum_annealing}{\bf Emulating the Transverse Ising Hamiltonian:} {\bf a.} {A 1D ferromagnetic linear chain ($J_{ij}$=$+2$) with $M = 8$ spins described by the transverse Ising Hamiltonian is emulated}. The average $z$ magnetization is shown as a function of the transverse magnetic field $\Gamma_x$ at an inverse temperature of $\beta = 20$ using a network of $8 \times 250$ p-bits with periodic boundary conditions {(shown as \textbf{b}). The 250 p-bits serve as replicas generated using a Suzuki-Trotter decomposition.} The exact quantum Boltzmann solution is compared against the average $z$ magnetization for different values of $s$. As $s$ is decreased, the system converges to the Boltzmann solution for a value of $s = 1/12$. As $s$ decreases beyond $s=1/12$, the system begins to diverge from the solution for larger $\Gamma_x/J$ due to the chosen precision of the digital implementation. For each $\Gamma_X$ and $s$, 200 samples from a free-running FPGA implementation were collected spaced $\approx30,000$ synapse delays apart. {\bf c.} Boxplots show the difference in computed mean at each $\Gamma_x/J$ from the ideal Boltzmann solution, for each value of $s$.}
 \label{fig:QA}
\end{figure}

Recently, it was theoretically argued \cite{camsari_scaled_2018} that a network of p-bits described by Eq.~\eqref{eq:synaptic_function} and interconnected according to Eq.~\eqref{eq:binary_stochastic_neuron} can be used to 
emulate a quantum system with a finite number of classical replicas, where a p-bit is represented in hardware by the stochastic MTJ-based implementation of Fig.~\ref{fig:apc:mtj_apc}. 

We start from the nearest neighbor Transverse Ising Hamiltonian in 1D \cite{pfeuty1970one}: 

\begin{equation}
 \mathcal{H}_{Q} \hspace{-3pt}=\hspace{-3pt}-\hspace{-3pt}\left(\sum_{i}^{M} J_{i,i+1} \sigma^z_i \sigma^z_{i+1} + \Gamma_x \sum_i^{M} \sigma^x_i + \Gamma_z \sum_i^{M}\sigma^z_i \right)
 \label{eq:isingH}
 \end{equation}
 where $J_{i,i+1}$ represents the interaction between neighboring spins, $\Gamma_x$ and $\Gamma_z$ are local magnetic fields in the $\hat{x}$ and $\hat{z}$ directions,{ $\sigma_i^{\{x,y,z\}}$ is the Pauli spin operator, and $M$ is the number of spins in the chain.} After a Suzuki-Trotter mapping,  the 2D classical Hamiltonian that approximates this system with $n$ replicas is given as: 
\begin{eqnarray}
\mathcal{H}_{C} &=& -\bigg(\sum_{k=1}^{n}\hspace{-1pt}\sum^{M}_{i=1} (J_{\parallel})_{i,i+1} \ m_{i,k} m_{i+1,k} + \gamma_z m_{i,k} \nonumber \\  
&& +J_{\perp} \  m_{i,k}m_{i, k+1}\bigg)
\label{eq:mappedH}
\end{eqnarray} 
where {the parallel coupling is} $(J_{\parallel})_{i,j}=J_{i,j}/n$, $\gamma_z  = \Gamma_z / n $ and the vertical coupling term is $J_{\perp}=-1/(2\beta)\mathrm{log \ tanh}(\beta \Gamma_x / n)$ and $m_{i,j} \in \{-1,+1\}$ {represent the classical spins in the 2D lattice.}

Eq.~\eqref{eq:mappedH} can be used to find each diagonal element of the quantum density matrix and therefore, all diagonal operators, and their correlations can be calculated from it. The corresponding interaction matrix ($J_{ij}$) to perform a p-bit simulation can be calculated from the mapped classical Hamiltonian, such that $I_i  = -\partial H_C/\partial m_i $ to yield the weight coefficients. Note that the Suzuki-Trotter decomposition adds another dimension to the classical system, therefore a 1D linear chain for a quantum system is emulated by a 2D classical system. 

In Fig.~\ref{fig:QA}, we simulate a ferromagnetic linear chain ($J_{i,j}=+2$) using $M=8$ spins with periodic boundary conditions at different transverse magnetic fields and we compute the average magnetization, $\langle m_z \rangle $. We include a symmetry breaking field in the $+$z direction ($\Gamma_z=+1)$ to obtain a net magnetization at vanishing transverse fields ($\Gamma_x = 0)$. We obtain the exact density matrix by directly diagonalizing the quantum Hamiltonian (Eq.~\eqref{eq:mappedH}): $\rho= \exp(-\beta\mathcal{H}_Q)$,  where we chose an inverse temperature of $\beta = 20$. Once $\rho$ is known, we compute the average magnetization by tracing it with the magnetization operator $S^z$, $\langle m \rangle = \mathrm{tr.}[S^z \rho]/\mathrm{tr.}[\rho]$. The exact solution is shown as a black solid line in Fig.~\ref{fig:QA:a}.  

The corresponding average magnetization is obtained by running the classical system that is described by Eq.~\eqref{eq:mappedH} with $n=250$ replicas (250 $\times$ 8 = $2000$ spins) using Eq.~\eqref{eq:autonomous_equations} in the FPGA emulator. Fig.~\ref{fig:QA:a} shows different $s$ values that are used to obtain the exact result. Clearly, choosing $s$ too high ($s=1/1$) fails, but gradually decreasing $s$ allows the result to approach the exact result. We observe that $s=1/12$ seems to be an optimal choice, as decreasing $s$ further does not yield more improvement due to the chosen precision in the digital implementation. Specifically, as $s$ continues to reduce, the chosen precision of the arithmetic logic and look-up tables in the FPGA emulator design limits the accuracy of the solution. {Fig.~\ref{fig:QA:c}} quantitatively shows a boxplot of the error incurred at each $s$ value across 200 trials. In the the Suzuki-Trotter decomposition, increasing $\Gamma_X/J$ systematically increases the error but a reasonably close agreement is observed for $s\leq 1/8$.

Typically, SQA algorithms initialize the system at a high magnetic field ($\Gamma_X)$ and slowly remove it to keep the system in its ground state and guide it to the desired ground state of the Ising Hamiltonian. In Fig.~\ref{fig:QA}, however, we have not changed the magnetic field as a function of time, but rather sampled from 200 ensembles, each separated by $\sim 30, 000$ synapse delays, to obtain the system statistics. This means that not only was the system guided to the expected ground state ($\Gamma_X\rightarrow 0, \langle m \rangle \rightarrow 1$), but it also followed the correct average magnetization at high magnetic fields. As such, this could be viewed as an example of sampling a probability distribution rather than finding the ground state of the system \cite{ackley_learning_1985}, an important problem space where an ApC could be useful.

It is important to note that to solve optimization problems using SQA, an exact mapping of the quantum system may not be optimal and a finite number of replicas that only approximate the thermodynamics of the quantum system could lead to more efficient results. In the present context, optimizing the number of replicas for a given system can be arranged since our system is \textit{not} a natural quantum system nor is it trying to faithfully reproduce the behavior of such a system. Therefore, optimizing the number of physical replicas or finding the replica with the lowest energy becomes possible in our engineered system \cite{heim2015quantum}.  

\begin{figure}[!t]
    \centering
    \setlength\abovecaptionskip{-0.5\baselineskip}
    \includegraphics[width=1\linewidth]{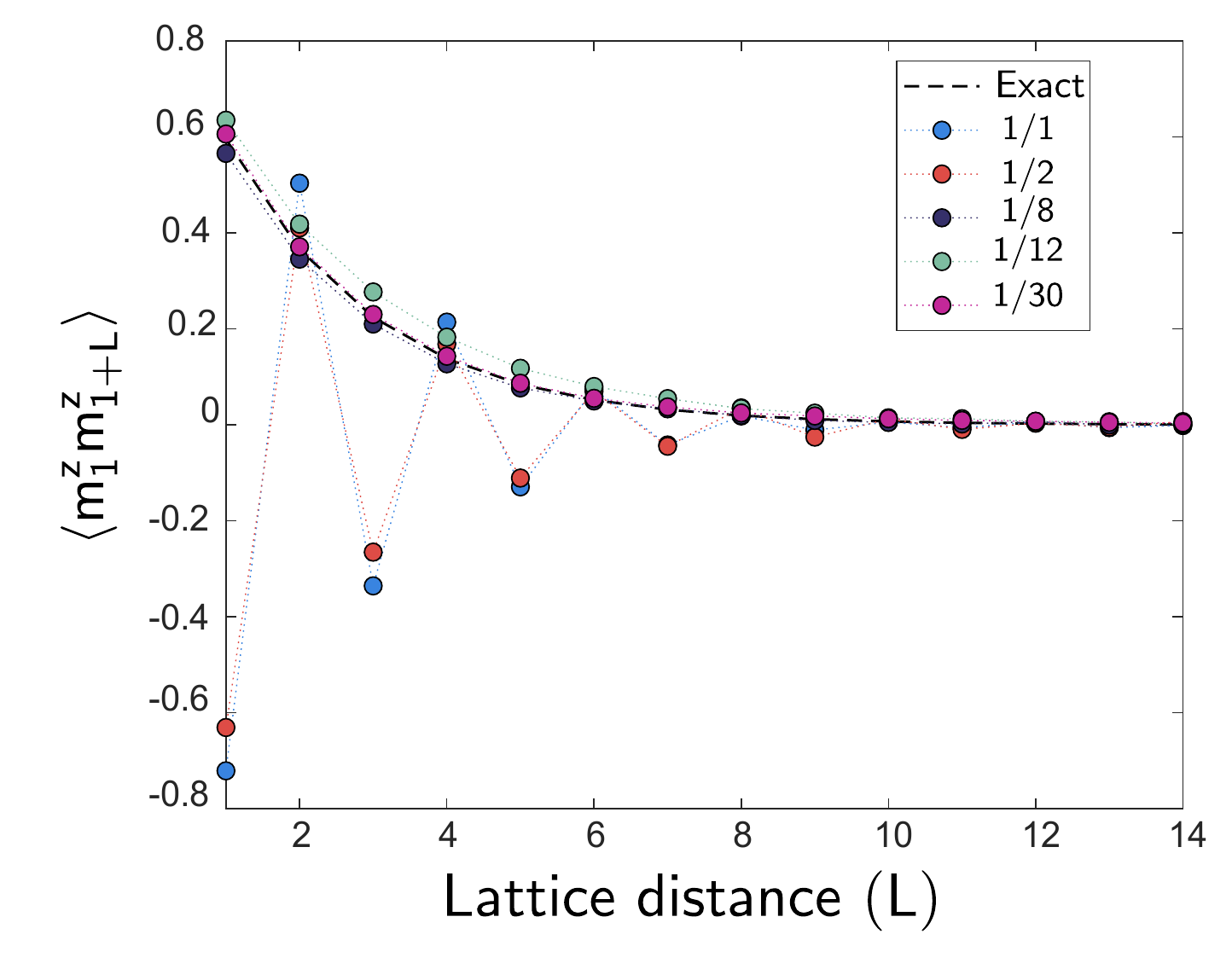}
    \caption{\label{fig:q250}{\bf Transverse Ising system with M=250 q-bits:} {\bf a.} A 1D ferromagnetic linear chain ($J_{ij}$=$+1$) with $M = 250$ spins is modeled using a network of 
        $10 \times 250$ p-bits with periodic boundary conditions at an inverse temperature of $\beta=0.744$ with $\Gamma_x=0.5$. The measured quantity is the average correlation measured between the first lattice point and different lattice distances ($L$), which decay to zero as $L$ increases. Results are shown at different $s$ values (inset). For each lattice distance and $s$, $10^5$ samples from a free-running FPGA implementation were collected spaced $\approx$640,000 synapse delays apart. Samples are further averaged over replicas to obtain the mean correlation for the quantum system.}
\end{figure}

{The replica discretization for the present implementation of SQA incurs errors of order O($\beta^3/r^2$) where $\beta$ is the inverse pseudo-temperature and $r$ is the number of replicas \cite{heim2015quantum}. The explicit form of the error provides a guide to reduce errors at a fixed $\beta$ at the expense of adding more p-bits. As an example, note the increasing error as $\Gamma_x/J$ in Fig.~\ref{fig:QA} where increasing $\Gamma_x$ compared to a fixed J is like increasing $\beta$.  In both examples (Fig.~\ref{fig:QA}-\ref{fig:q250}) we discuss, we have made comparisons to \textit{exact} calculations that did not require a careful optimization of the number of replicas since we were guided by the errors with respect to exact calculations.}

As a second example, we illustrate the trade-off between the number of replicas and the size of the quantum system when limited to a fixed number of p-bits, for example in an FPGA with finite resources. The emulation in Fig.~\ref{fig:quantum_annealing} modeled a 1D Transverse Field Ising Model (TFIM) with $M=8$ spins and with 250 replicas (with a total of 2000 p-bits) to obtain an accurate result at a low (dimensionless) temperature of $\beta=20$. In Fig.~\ref{fig:q250}, we show another example with a much larger  system of $M=250$ spins but using only 10 replicas per quantum spin for a total of 2500 p-bits. 

In this example, instead of plotting average magnetization, we plot the average correlation (i.e, $\langle m^z_1 m^z_{1+L}\rangle)$ between lattice points at a fixed inverse temperature ($\beta=0.744$) and a given magnetic field $\Gamma=0.5$. We compare the FPGA results with an exact calculation that is known for the 1D TFIM model at a finite magnetic field. Unlike the classical Ising model where such correlations can be easily computed, the exact calculation for the corresponding quantum problem is quite involved and we omit the details here. In short, the correlation for a finite sized 1D lattice at a given $\Gamma$ and $\beta$ can be computed exactly by constructing a Toeplitz determinant, $T_n$, where $n$ represents the correlation value at a maximum distance from a reference lattice point that is arbitrarily chosen to be $1$ in the present example (See Chapter 10, Equation (10.58) and the preceding discussion in Ref.\cite{sachdev_2011}). The results that we obtain from this exact method and the FPGA sampling are shown in Fig.~\ref{fig:q250} where we observe excellent agreement between the ApC results and the exact calculation even for a relatively low number of replicas ($r=10$) when the $s$ value is less than or equal to $ 1/8$ which corresponds to approximately $\approx 1/8\times 2500\approx 312$ simultaneous updates. When the $s$ parameter exceeds this value, we observe systematic oscillations and large errors in the average correlation arising from a large degree of parallel updates, reminiscent of similar oscillations that are due to simultaneous parallel updating in Boltzmann networks\cite{hinton_coursera}.

\section{Discussion}
\label{sec:discussion}
\begin{table*}[]
    \begin{center}
    \begin{threeparttable}
        \begin{tabular}{c|cc|ccc |}
            
            & \multicolumn{2}{c|}{\textbf{Sequenced}} & \multicolumn{3}{c|}{\textbf{Autonomous (This Work)}} \\ \cline{2-6} \\  [-7pt] 
            & \multicolumn{1}{c|}{\textbf{Hitachi\tnote{a}}} & \multicolumn{1}{c|}{\textbf{Janus II}\tnote{b}} & \multicolumn{1}{c|}{\textbf{2K QA}\tnote{c}} & \multicolumn{1}{c|}{\textbf{8K  Ising}\tnote{c}} & \multicolumn{1}{c|}{\textbf{Projected}\tnote{d}} \\
            \cline{2-6}\\  [-7pt] 
             \textbf{Technology} & CMOS (SRAM) & CMOS (FPGA) & CMOS (FPGA) & CMOS (FPGA) & CMOS + MTJ \\
            \textbf{Total Power (W)} & 0.05 & 25 & 55 & 32 & 19.25 \\ 
            \textbf{Number of Neurons (N)} & 20K ($80 \times 256$) & 2K & 2K ($8\times 250$) & 8.1K ($90 \times 90$) & 1M \\
            \textbf{$s \equiv \tau_{S}/\tau_{N}$} & $1^\dagger$ & $1^\dagger$ & 1/12 & 1/4 & 1/10 \\
            \textbf{Synapse Delay ($\tau_S$)} (ps) & 10,000 & 4,000 & 8,000 & 8,000 & 10 \\
            \textbf{Neuron Delay ($\tau_N$)} (ps) & 10,000 & 4,000 & 96,000 & 32,000 & 100 \\
            \textbf{Flips per Second ($f$)} & $1\times 10^{12}$ & $2.5\times 10^{11}$ & $2.08\times 10^{10}$ & $2.5\times 10^{11}$ & $\boxed{1\times 10^{16}}$ \\
            \textbf{Spin Update Time ($1/f$) (ps/flip)} & 1 & 4 & 48 & 4 & \boxed{\textbf{0.0001}} \\
            \textbf{Energy per Flip ($\epsilon$) (nJ/flip)} & $5 \times 10^{-5}$ & 0.1 & 2.64 & 0.13 & $ 1.93\times 10^{-6}$ \\
            \bottomrule\addlinespace[1ex]          
        \end{tabular}
        \begin{tablenotes}\footnotesize
            \item[a] Calculations based on information extracted from \cite{yamaoka_20k-spin_2016}. Design used sequential implementation, therefore we assume ($\dagger$) $s = 1$ and an optimal updating scheme such that half of all neurons are updated every step such that $f = (0.5)N/\tau_S$. Synapse and neuron delays assumed to operate at interaction frequency of 100 MHz.
            \item[b] Calculations based on information extracted from \cite{baity-jesi_janus_2014}. Janus II supports a much larger number of neurons using external memory, multiple FPGAs, and data shuttling between devices. Here we limit the comparison to only a single spin processor assuming all spins are co-located on one FPGA, as would be required for an autonomous design. As with (a), we assume ($\dagger$) $s = 1$. With all spins on one chip, an optimal updating scheme updates half of all neurons every step for $f = (0.5)N/\tau_S$. Based on this, the spin update time calculated here is 4 instead of 2 as in \cite{baity-jesi_janus_2014}. {We note that this discrepancy is negligible when compared to the projected autonomous coprocessor that can be orders of magnitude faster.} Design assumed to operate at operating frequency of 250 MHz.
            \item[c] Power consumption measured from maximum power draw during computation. 2K QA topology based on 16-bit precision operations and 8K Ising is based on 2-bit precision operations. Selected $s$ values based on Figs.~\ref{fig:max_cut} and \ref{fig:quantum_annealing}. 
            \item[d] Assuming chip density of 1M based on memory density available from commercial ST-MRAM \cite{rizzo_fully_2013}. With nearest-neighbor connections, a synapse delay of 10 ps assumed \cite{peng_gu_technological_2015}. Assumed neuron delay of $\sim$100 ps and a steady-state neuron power consumption 10 $\mu$W \cite{hassan_low-barrier_2019}. Overhead from nearest-neighbor memresistive cross-bar and external communication logic estimated as $\sim 9.25$ W using 5 memresistors and 1 op-amp per neuron \cite{li_memristor-based_2013}, similar order of magnitude values can be obtained from \cite{cai2019harnessing} .
        \end{tablenotes}
    \end{threeparttable}
    
    \caption{\label{tab:comparison} {\bf Comparison of Sequential and Asynchronous Ising Computers:} Ising computers proposed to-date have used sequential updating mechanisms based on CMOS technology. These designs achieve spin update times of $\sim 1-4$ ps/flip. The 20K SRAM chip from \cite{yamaoka_20k-spin_2016} achieves an energy of 50 fJ per flip. The autonomous CMOS implementations demonstrated in this work obtain a similar spin update time as sequenced implementations and comparative energy per flip as sequential FPGA implementations. Using CMOS and MTJ technologies as proposed herein, a highly efficient design with petaflips per second with 2 fJ per flip is projected.
    }
    \end{center}
\end{table*}

These example applications help demonstrate the feasibility of an ApC governed by Eq.~\eqref{eq:autonomous_equations} to follow Boltzmann statistics without the need for a controlling sequencer. But in order to make use of ultrafast flip rates, \textit{f}, enabled by short $\tau_S$ times, it is important that the building blocks be energy efficient to ensure that power levels are acceptable: 
\begin{equation}
P = f \ \varepsilon
\end{equation}
where $P$ is the power budget and $\varepsilon$ is the energy required per flip.

In Table~\ref{tab:comparison}, we compare representative Ising computers based on sequenced designs \cite{yamaoka_20k-spin_2016,baity-jesi_janus_2014} to the autonomous designs implemented in and projected by this work, focusing on nearest-neighbor implementations for this discussion. The last row of the table shows $\varepsilon$ for both the sequenced and autonomous designs. As shown in the table, the FPGA based 8K-spin ApC achieves an energy per flip that is similar to the Janus II sequenced FPGA implementation. However, this comparison applies some artificial constraints on the Janus II design: namely that all spins must be simultaneously resident in the device at one time. This is a requirement for unclocked autonomous designs that for the sake of comparison we applied to the sequenced design. A clocked, sequenced design can pause the system and leverage external memory for storage of neuron state, providing a large pool of logical neurons, though limited by available memory bandwidth. Additionally, the ability to pause the network enables time-multiplexing and can harness the ability to re-use logic resources.

The FPGA results naturally consume more power and have reduced density\cite{kuon_measuring_2007}. The 8K spin result of Table \ref{tab:comparison} has $\sim 18$ W of static power dissipation due to the periphery included in the FPGA design, not all of which is used. Using the FPGA to ASIC power ratio of 14\cite{kuon_measuring_2007}, a naive migration of the design to an ASIC would result in an $\varepsilon$ of $\sim 4.0\times 10^{-3}$. By translating these approaches to a tailored ASIC implementation, the energy efficiency per flip increases  and the ability to increase the density of resident neurons within a device increases substantially. Extending this further, it should be feasible to obtain designs with  $\sim 10$ \textit{petaflips per second} with a power budget of $\sim 10$ W, but we need devices with $\varepsilon \sim 1$ fJ that also support a density of 1M devices.

The CMOS based SRAM design of Hitachi\cite{yamaoka_20k-spin_2016} has an estimated energy per flip of $\varepsilon \sim 50$ fJ, though the neuron density limits the ability of the approach to obtain an $f$ of petaflips per second. Using a hybrid CMOS and MTJ design as would be encountered in modern commercial MRAMs \cite{rizzo_fully_2013}, it should be feasible to obtain $\sim 10 $ petaflips per sec as projected in the last column of Table~\ref{tab:comparison} \cite{hassan_low-barrier_2019}. Modern MRAMs can achieve Gb densities; however, we limit the projection to a 1 Mb density based on a target power consumption of $\sim 20$ W for the neurons and synapses in the design.

It should be noted that the 20 W power target was arbitrarily chosen. In principle the autonomous designs can leverage clocking to choose when global p-bit updates should occur, much like the FPGA emulator discussed in the methods section. While this approach can save power, it limits the utility of the MTJ based approach by constraining the flips per second ($f$) to the same limits of Eq.~\eqref{eq:flips_per_second_sequenced} due to the clocking scheme. Even with this limit in place, as the connectivity between neurons increases beyond nearest-neighbor, the sequencing logic becomes more complex while the ApC only requires a balance of $s$ to ensure proper convergence, though both approaches still face challenges with routing congestion as $N$ grows. In the case of all-to-all connectivity, the sequencing logic reduces in complexity and technically can be implemented with a single time-multiplexed weighted p-bit. In this situation, the benefits of an ApC begin to degrade, except for the elimination of memory bottlenecks with the use of distributed weights and the avoidance of time-multiplexing. A 500 node all-to-all network was implemented using the FPGA emulator, and the resulting $s$ was directly proportional to the number of neurons, affirming the reduced benefit.

\section{Conclusion and Future Work}
\label{sec:conclusion}

{In this work we presented a vision for an autonomous probabilistic computer for applications in optimization and machine learning. Using stochastic nanomagnets as intrinsic p-bits, an ApC based on these devices provides an opportunity to realize a scalable co-processor achieving high-performance operation. As the technology is not yet available to build a scaled device, we presented a benchmarked behavior model for the autonomous computer that facilitated direct study of the ApC. This behavioral model was implemented in an FPGA to establish a framework for the study of various applications and design trade-offs. The results presented in Table \ref{tab:comparison} demonstrate that with the removal of sequencers, the ability to run at speeds limited only by synapse delays, and the ability scale to millions of neurons, all within an accessible power budget, the proposed ApC is a compelling alternative to clocked, sequential designs for stochastic ANN coprocessing. 

We have also emphasized a key metric, \textit{flips per second}, as a figure-of-merit for emerging probabilistic annealers that quantify their performance in terms of a problem or substrate independent manner. Improving flips per second by application specific, massively parallel or autonomous architectures using nanodevices as we have suggested could lead to efficient domain-specific, probabilistic coprocessors that can outperform conventional implementations in the future.}



\begin{thebibliography}{10}
    	\providecommand{\url}[1]{#1}
    	\csname url@samestyle\endcsname
    	\providecommand{\newblock}{\relax}
    	\providecommand{\bibinfo}[2]{#2}
    	\providecommand{\BIBentrySTDinterwordspacing}{\spaceskip=0pt\relax}
    	\providecommand{\BIBentryALTinterwordstretchfactor}{4}
    	\providecommand{\BIBentryALTinterwordspacing}{\spaceskip=\fontdimen2\font plus
    		\BIBentryALTinterwordstretchfactor\fontdimen3\font minus
    		\fontdimen4\font\relax}
    	\providecommand{\BIBforeignlanguage}[2]{{%
    			\expandafter\ifx\csname l@#1\endcsname\relax
    			\typeout{** WARNING: IEEEtran.bst: No hyphenation pattern has been}%
    			\typeout{** loaded for the language `#1'. Using the pattern for}%
    			\typeout{** the default language instead.}%
    			\else
    			\language=\csname l@#1\endcsname
    			\fi
    			#2}}
    	\providecommand{\BIBdecl}{\relax}
    	\BIBdecl
    	
    	\bibitem{schuman_survey_2017}
    	C.~D. Schuman, T.~E. Potok, R.~M. Patton, J.~D. Birdwell, M.~E. Dean, G.~S.
    	Rose, and J.~S. Plank, ``A {Survey} of {Neuromorphic} {Computing} and
    	{Neural} {Networks} in {Hardware},'' \emph{arXiv:1705.06963 [cs]}, May 2017,
    	arXiv: 1705.06963.
    	
    	\bibitem{detorakis2019inherent}
    	G.~Detorakis, S.~Dutta, A.~Khanna, M.~Jerry, S.~Datta, and E.~Neftci,
    	``Inherent weight normalization in stochastic neural networks,'' in
    	\emph{Advances in Neural Information Processing Systems}, 2019, pp.
    	3286--3297.
    	
    	\bibitem{buesing2011neural}
    	L.~Buesing, J.~Bill, B.~Nessler, and W.~Maass, ``Neural dynamics as sampling: a
    	model for stochastic computation in recurrent networks of spiking neurons,''
    	\emph{PLoS computational biology}, vol.~7, no.~11, p. e1002211, 2011.
    	
    	\bibitem{courbariaux2016binarized}
    	M.~Courbariaux, I.~Hubara, D.~Soudry, R.~El-Yaniv, and Y.~Bengio, ``Binarized
    	neural networks: Training deep neural networks with weights and activations
    	constrained to+ 1 or-1,'' \emph{arXiv preprint arXiv:1602.02830}, 2016.
    	
    	\bibitem{boixo_evidence_2014}
    	S.~Boixo, T.~F. R{\o}nnow, S.~V. Isakov, Z.~Wang, D.~Wecker, D.~A. Lidar, J.~M.
    	Martinis, and M.~Troyer, ``Evidence for quantum annealing with more than one
    	hundred qubits,'' \emph{Nature Physics}, vol.~10, no.~3, pp. 218--224, Mar.
    	2014.
    	
    	\bibitem{yamaoka_20k-spin_2016}
    	M.~Yamaoka, C.~Yoshimura, M.~Hayashi, T.~Okuyama, H.~Aoki, and H.~Mizuno, ``A
    	20k-{Spin} {Ising} {Chip} to {Solve} {Combinatorial} {Optimization}
    	{Problems} {With} {CMOS} {Annealing},'' \emph{IEEE Journal of Solid-State
    		Circuits}, vol.~51, no.~1, pp. 303--309, Jan. 2016.
    	
    	\bibitem{mcmahon_fully-programmable_2016}
    	P.~L. McMahon, A.~Marandi, Y.~Haribara, R.~Hamerly, C.~Langrock, S.~Tamate,
    	T.~Inagaki, H.~Takesue, S.~Utsunomiya, K.~Aihara, R.~L. Byer, M.~M. Fejer,
    	H.~Mabuchi, and Y.~Yamamoto, ``A fully-programmable 100-spin coherent {Ising}
    	machine with all-to-all connections,'' \emph{Science}, p. aah5178, Oct. 2016.
    	
    	\bibitem{inagaki_coherent_2016}
    	T.~Inagaki, Y.~Haribara, K.~Igarashi, T.~Sonobe, S.~Tamate, T.~Honjo,
    	A.~Marandi, P.~L. McMahon, T.~Umeki, K.~Enbutsu, O.~Tadanaga, H.~Takenouchi,
    	K.~Aihara, K.-i. Kawarabayashi, K.~Inoue, S.~Utsunomiya, and H.~Takesue, ``A
    	coherent {Ising} machine for 2000-node optimization problems,''
    	\emph{Science}, vol. 354, no. 6312, pp. 603--606, Nov. 2016.
    	
    	\bibitem{okuyama_ising_2017}
    	T.~Okuyama, M.~Hayashi, and M.~Yamaoka, ``An {Ising} {Computer} {Based} on
    	{Simulated} {Quantum} {Annealing} by {Path} {Integral} {Monte} {Carlo}
    	{Method},'' in \emph{2017 {IEEE} {International} {Conference} on {Rebooting}
    		{Computing} ({ICRC})}, Nov. 2017, pp. 1--6.
    	
    	\bibitem{sutton_intrinsic_2017}
    	B.~Sutton, K.~Y. Camsari, B.~Behin-Aein, and S.~Datta, ``Intrinsic optimization
    	using stochastic nanomagnets,'' \emph{Scientific Reports}, vol.~7, p. 44370,
    	Mar. 2017.
    	
    	\bibitem{hamerly_experimental_2019}
    	R.~Hamerly, T.~Inagaki, P.~L. McMahon, D.~Venturelli, A.~Marandi, T.~Onodera,
    	E.~Ng, C.~Langrock, K.~Inaba, T.~Honjo, K.~Enbutsu, T.~Umeki, R.~Kasahara,
    	S.~Utsunomiya, S.~Kako, K.-i. Kawarabayashi, R.~L. Byer, M.~M. Fejer,
    	H.~Mabuchi, D.~Englund, E.~Rieffel, H.~Takesue, and Y.~Yamamoto,
    	``\BIBforeignlanguage{en}{Experimental investigation of performance
    		differences between coherent {Ising} machines and a quantum annealer},''
    	\emph{\BIBforeignlanguage{en}{Science Advances}}, vol.~5, no.~5, p. eaau0823,
    	May 2019.
    	
    	\bibitem{wang_oscillator-based_2017}
    	T.~Wang and J.~Roychowdhury, ``Oscillator-based {Ising} {Machine},''
    	\emph{arXiv:1709.08102 [physics]}, Sep. 2017, arXiv: 1709.08102.
    	
    	\bibitem{wang_oim:_2019}
    	------, ``\BIBforeignlanguage{en}{{OIM}: {Oscillator}-{Based} {Ising}
    		{Machines} for {Solving} {Combinatorial} {Optimisation} {Problems}},'' in
    	\emph{\BIBforeignlanguage{en}{Unconventional {Computation} and {Natural}
    			{Computation}}}, ser. Lecture {Notes} in {Computer} {Science}, I.~McQuillan
    	and S.~Seki, Eds.\hskip 1em plus 0.5em minus 0.4em\relax Springer
    	International Publishing, 2019, pp. 232--256.
    	
    	\bibitem{raychowdhury_computing_2019}
    	A.~Raychowdhury, A.~Parihar, G.~H. Smith, V.~Narayanan, G.~Csaba, M.~Jerry,
    	W.~Porod, and S.~Datta, ``Computing {With} {Networks} of {Oscillatory}
    	{Dynamical} {Systems},'' \emph{Proceedings of the IEEE}, vol. 107, no.~1, pp.
    	73--89, Jan. 2019.
    	
    	\bibitem{aramon_physics_2019}
    	\BIBentryALTinterwordspacing
    	M.~Aramon, G.~Rosenberg, E.~Valiante, T.~Miyazawa, H.~Tamura, and H.~G.
    	Katzgraber, ``Physics-inspired optimization for quadratic unconstrained
    	problems using a digital annealer,'' \emph{Frontiers in Physics}, vol.~7,
    	p.~48, 2019. [Online]. Available:
    	\url{https://www.frontiersin.org/article/10.3389/fphy.2019.00048}
    	\BIBentrySTDinterwordspacing
    	
    	\bibitem{goto_combinatorial_2019}
    	\BIBentryALTinterwordspacing
    	H.~Goto, K.~Tatsumura, and A.~R. Dixon, ``Combinatorial optimization by
    	simulating adiabatic bifurcations in nonlinear hamiltonian systems,''
    	\emph{Science Advances}, vol.~5, no.~4, 2019. [Online]. Available:
    	\url{https://advances.sciencemag.org/content/5/4/eaav2372}
    	\BIBentrySTDinterwordspacing
    	
    	\bibitem{yamamoto_statica_2020}
    	K.~{Yamamoto}, K.~{Ando}, N.~{Mertig}, T.~{Takemoto}, M.~{Yamaoka},
    	H.~{Teramoto}, A.~{Sakai}, S.~{Takamaeda-Yamazaki}, and M.~{Motomura}, ``7.3
    	statica: A 512-spin 0.25m-weight full-digital annealing processor with a
    	near-memory all-spin-updates-at-once architecture for combinatorial
    	optimization with complete spin-spin interactions,'' in \emph{2020 IEEE
    		International Solid- State Circuits Conference - (ISSCC)}, 2020, pp.
    	138--140.
    	
    	\bibitem{su_cim_2020}
    	Y.~{Su}, H.~{Kim}, and B.~{Kim}, ``31.2 cim-spin: A 0.5-to-1.2v scalable
    	annealing processor using digital compute-in-memory spin operators and
    	register-based spins for combinatorial optimization problems,'' in \emph{2020
    		IEEE International Solid- State Circuits Conference - (ISSCC)}, 2020, pp.
    	480--482.
    	
    	\bibitem{lucas_ising_2014}
    	A.~Lucas, ``Ising formulations of many {NP} problems,'' \emph{Interdisciplinary
    		Physics}, vol.~2, p.~5, 2014.
    	
    	\bibitem{dattani2019quadratization}
    	N.~Dattani, ``Quadratization in discrete optimization and quantum mechanics,''
    	\emph{arXiv preprint arXiv:1901.04405}, 2019.
    	
    	\bibitem{ackley_learning_1985}
    	D.~H. Ackley, G.~E. Hinton, and T.~J. Sejnowski, ``A {Learning} {Algorithm} for
    	{Boltzmann} {Machines}*,'' \emph{Cognitive Science}, vol.~9, no.~1, pp.
    	147--169, Jan. 1985.
    	
    	\bibitem{neal_connectionist_1992}
    	R.~M. Neal, ``Connectionist learning of belief networks,'' \emph{Artificial
    		Intelligence}, vol.~56, no.~1, pp. 71--113, Jul. 1992.
    	
    	\bibitem{aarts_simulated_1989}
    	E.~Aarts and J.~Korst, \emph{Simulated {Annealing} and {Boltzmann} {Machines}:
    		{A} {Stochastic} {Approach} to {Combinatorial} {Optimization} and {Neural}
    		{Computing}}.\hskip 1em plus 0.5em minus 0.4em\relax New York, NY, USA: John
    	Wiley \& Sons, Inc., 1989.
    	
    	\bibitem{haykin2009neural}
    	S.~S. Haykin \emph{et~al.}, \emph{Neural networks and learning machines/Simon
    		Haykin.}\hskip 1em plus 0.5em minus 0.4em\relax New York: Prentice Hall,,
    	2009.
    	
    	\bibitem{camsari_stochastic_2017}
    	K.~Y. Camsari, R.~Faria, B.~M. Sutton, and S.~Datta, ``Stochastic p -{Bits} for
    	{Invertible} {Logic},'' \emph{Physical Review X}, vol.~7, no.~3, Jul. 2017.
    	
    	\bibitem{borders2019integer}
    	W.~A. Borders, A.~Z. Pervaiz, S.~Fukami, K.~Y. Camsari, H.~Ohno, and S.~Datta,
    	``Integer factorization using stochastic magnetic tunnel junctions,''
    	\emph{Nature}, vol. 573, no. 7774, pp. 390--393, 2019.
    	
    	\bibitem{hassan_low-barrier_2019}
    	O.~Hassan, R.~Faria, K.~Y. Camsari, J.~Z. Sun, and S.~Datta, ``Low-{Barrier}
    	{Magnet} {Design} for {Efficient} {Hardware} {Binary} {Stochastic}
    	{Neurons},'' \emph{IEEE Magnetics Letters}, vol.~10, pp. 1--5, 2019.
    	
    	\bibitem{feynman_simulating_1982}
    	R.~P. Feynman, ``Simulating physics with computers,'' \emph{International
    		Journal of Theoretical Physics}, vol.~21, no. 6-7, pp. 467--488, Jun. 1982.
    	
    	\bibitem{merolla_million_2014}
    	P.~A. Merolla, J.~V. Arthur, R.~Alvarez-Icaza, A.~S. Cassidy, J.~Sawada,
    	F.~Akopyan, B.~L. Jackson, N.~Imam, C.~Guo, Y.~Nakamura, B.~Brezzo, I.~Vo,
    	S.~K. Esser, R.~Appuswamy, B.~Taba, A.~Amir, M.~D. Flickner, W.~P. Risk,
    	R.~Manohar, and D.~S. Modha, ``\BIBforeignlanguage{en}{A million
    		spiking-neuron integrated circuit with a scalable communication network and
    		interface},'' \emph{\BIBforeignlanguage{en}{Science}}, vol. 345, no. 6197,
    	pp. 668--673, Aug. 2014.
    	
    	\bibitem{davies_loihi:_2018}
    	M.~Davies, N.~Srinivasa, T.~Lin, G.~Chinya, Y.~Cao, S.~H. Choday, G.~Dimou,
    	P.~Joshi, N.~Imam, S.~Jain, Y.~Liao, C.~Lin, A.~Lines, R.~Liu,
    	D.~Mathaikutty, S.~McCoy, A.~Paul, J.~Tse, G.~Venkataramanan, Y.~Weng,
    	A.~Wild, Y.~Yang, and H.~Wang, ``Loihi: {A} {Neuromorphic} {Manycore}
    	{Processor} with {On}-{Chip} {Learning},'' \emph{IEEE Micro}, vol.~38, no.~1,
    	pp. 82--99, Jan. 2018.
    	
    	\bibitem{fukushima2014spin}
    	A.~Fukushima, T.~Seki, K.~Yakushiji, H.~Kubota, H.~Imamura, S.~Yuasa, and
    	K.~Ando, ``Spin dice: A scalable truly random number generator based on
    	spintronics,'' \emph{Applied Physics Express}, vol.~7, no.~8, p. 083001,
    	2014.
    	
    	\bibitem{grollier_spintronic_2016}
    	J.~Grollier, D.~Querlioz, and M.~D. Stiles, ``Spintronic {Nanodevices} for
    	{Bioinspired} {Computing},'' \emph{Proceedings of the IEEE}, vol. 104,
    	no.~10, pp. 2024--2039, Oct. 2016.
    	
    	\bibitem{behin-aein_building_2016}
    	B.~Behin-Aein, V.~Diep, and S.~Datta, ``A building block for hardware belief
    	networks,'' \emph{Scientific Reports}, vol.~6, p. 29893, Jul. 2016.
    	
    	\bibitem{liyanagedera_stochastic_2017}
    	C.~M. Liyanagedera, A.~Sengupta, A.~Jaiswal, and K.~Roy, ``Stochastic {Spiking}
    	{Neural} {Networks} {Enabled} by {Magnetic} {Tunnel} {Junctions}: {From}
    	{Nontelegraphic} to {Telegraphic} {Switching} {Regimes},'' \emph{Physical
    		Review Applied}, vol.~8, no.~6, p. 064017, Dec. 2017.
    	
    	\bibitem{parks2018superparamagnetic}
    	B.~Parks, M.~Bapna, J.~Igbokwe, H.~Almasi, W.~Wang, and S.~A. Majetich,
    	``Superparamagnetic perpendicular magnetic tunnel junctions for true random
    	number generators,'' \emph{AIP Advances}, vol.~8, no.~5, p. 055903, 2018.
    	
    	\bibitem{vodenicarevic2017low}
    	D.~Vodenicarevic, N.~Locatelli, A.~Mizrahi, J.~S. Friedman, A.~F. Vincent,
    	M.~Romera, A.~Fukushima, K.~Yakushiji, H.~Kubota, S.~Yuasa \emph{et~al.},
    	``Low-energy truly random number generation with superparamagnetic tunnel
    	junctions for unconventional computing,'' \emph{Physical Review Applied},
    	vol.~8, no.~5, p. 054045, 2017.
    	
    	\bibitem{rangarajan2017energy}
    	N.~Rangarajan, A.~Parthasarathy, N.~Kani, and S.~Rakheja, ``Energy-efficient
    	computing with probabilistic magnetic bits---performance modeling and
    	comparison against probabilistic cmos logic,'' \emph{IEEE Transactions on
    		Magnetics}, vol.~53, no.~11, pp. 1--10, 2017.
    	
    	\bibitem{lv_single_2017}
    	Y.~Lv and J.~Wang, ``A single magnetic-tunnel-junction stochastic computing
    	unit,'' in \emph{2017 {IEEE} {International} {Electron} {Devices} {Meeting}
    		({IEDM})}, Dec. 2017, pp. 36.2.1--36.2.4.
    	
    	\bibitem{mizrahi_neural-like_2018}
    	A.~Mizrahi, T.~Hirtzlin, A.~Fukushima, H.~Kubota, S.~Yuasa, J.~Grollier, and
    	D.~Querlioz, ``\BIBforeignlanguage{En}{Neural-like computing with populations
    		of superparamagnetic basis functions},'' \emph{\BIBforeignlanguage{En}{Nature
    			Communications}}, vol.~9, no.~1, p. 1533, Apr. 2018.
    	
    	\bibitem{camsari_p-bits_2019}
    	K.~Y. Camsari, B.~M. Sutton, and S.~Datta, ``p-bits for probabilistic spin
    	logic,'' \emph{Applied Physics Reviews}, vol.~6, no.~1, p. 011305, Mar. 2019.
    	
    	\bibitem{zand_composable_2019}
    	R.~Zand, K.~Y. Camsari, S.~Datta, and R.~F. Demara, ``Composable
    	{Probabilistic} {Inference} {Networks} {Using} {MRAM}-based {Stochastic}
    	{Neurons},'' \emph{J. Emerg. Technol. Comput. Syst.}, vol.~15, no.~2, pp.
    	17:1--17:22, Mar. 2019.
    	
    	\bibitem{nasrin2019low}
    	S.~Nasrin, J.~L. Drobitch, S.~Bandyopadhyay, and A.~R. Trivedi, ``Low power
    	restricted boltzmann machine using mixed-mode magneto-tunneling junctions,''
    	\emph{IEEE Electron Device Letters}, vol.~40, no.~2, pp. 345--348, 2019.
    	
    	\bibitem{daniels2019energy}
    	M.~W. Daniels, A.~Madhavan, P.~Talatchian, A.~Mizrahi, and M.~D. Stiles,
    	``Energy-efficient stochastic computing with superparamagnetic tunnel
    	junctions,'' \emph{arXiv preprint arXiv:1911.11204}, 2019.
    	
    	\bibitem{drobitch2019reliability}
    	J.~L. Drobitch and S.~Bandyopadhyay, ``Reliability and scalability of p-bits
    	implemented with low energy barrier nanomagnets,'' \emph{IEEE Magnetics
    		Letters}, vol.~10, pp. 1--4, 2019.
    	
    	\bibitem{lv2019experimental}
    	Y.~Lv, R.~P. Bloom, and J.-P. Wang, ``Experimental demonstration of
    	probabilistic spin logic by magnetic tunnel junctions,'' \emph{IEEE Magnetics
    		Letters}, 2019.
    	
    	\bibitem{camsari_implementing_2017}
    	K.~Y. Camsari, S.~Salahuddin, and S.~Datta, ``Implementing p-bits {With}
    	{Embedded} {MTJ},'' \emph{IEEE Electron Device Letters}, vol.~38, no.~12, pp.
    	1767--1770, Dec. 2017.
    	
    	\bibitem{calhoun_digital_2008}
    	B.~H. Calhoun, Y.~Cao, X.~Li, K.~Mai, L.~T. Pileggi, R.~A. Rutenbar, and K.~L.
    	Shepard, ``Digital {Circuit} {Design} {Challenges} and {Opportunities} in the
    	{Era} of {Nanoscale} {CMOS},'' \emph{Proceedings of the IEEE}, vol.~96,
    	no.~2, pp. 343--365, Feb. 2008.
    	
    	\bibitem{peng_gu_technological_2015}
    	{Peng Gu}, {Boxun Li}, {Tianqi Tang}, S.~Yu, {Yu Cao}, Y.~Wang, and H.~Yang,
    	``Technological exploration of {RRAM} crossbar array for matrix-vector
    	multiplication,'' in \emph{The 20th {Asia} and {South} {Pacific} {Design}
    		{Automation} {Conference}}, Jan. 2015, pp. 106--111.
    	
    	\bibitem{pervaiz_hardware_2017}
    	A.~Z. Pervaiz, L.~A. Ghantasala, K.~Y. Camsari, and S.~Datta, ``Hardware
    	emulation of stochastic p-bits for invertible logic,'' \emph{Scientific
    		Reports}, vol.~7, p. 10994, Sep. 2017.
    	
    	\bibitem{choi_minor-embedding_2008}
    	V.~Choi, ``Minor-embedding in adiabatic quantum computation: {I}. {The}
    	parameter setting problem,'' \emph{Quantum Information Processing}, vol.~7,
    	no.~5, pp. 193--209, Oct. 2008.
    	
    	\bibitem{choi_minor-embedding_2011}
    	------, ``Minor-embedding in adiabatic quantum computation: {II}.
    	{Minor}-universal graph design,'' \emph{Quantum Information Processing},
    	vol.~10, no.~3, pp. 343--353, Jun. 2011.
    	
    	\bibitem{butler_switching_2012}
    	W.~H. {Butler}, T.~{Mewes}, C.~K.~A. {Mewes}, P.~B. {Visscher}, W.~H.
    	{Rippard}, S.~E. {Russek}, and R.~{Heindl}, ``Switching distributions for
    	perpendicular spin-torque devices within the macrospin approximation,''
    	\emph{IEEE Transactions on Magnetics}, vol.~48, no.~12, pp. 4684--4700, Dec
    	2012.
    	
    	\bibitem{torunbalci2018modular}
    	M.~M. Torunbalci, P.~Upadhyaya, S.~A. Bhave, and K.~Y. Camsari, ``Modular
    	compact modeling of mtj devices,'' \emph{IEEE Transactions on Electron
    		Devices}, vol.~65, no.~10, pp. 4628--4634, 2018.
    	
    	\bibitem{behin2010proposal}
    	B.~Behin-Aein, D.~Datta, S.~Salahuddin, and S.~Datta, ``Proposal for an
    	all-spin logic device with built-in memory,'' \emph{Nature nanotechnology},
    	vol.~5, no.~4, p. 266, 2010.
    	
    	\bibitem{camsari2017implementing}
    	K.~Y. Camsari, S.~Salahuddin, and S.~Datta, ``Implementing p-bits with embedded
    	mtj,'' \emph{IEEE Electron Device Letters}, vol.~38, no.~12, pp. 1767--1770,
    	2017.
    	
    	\bibitem{sherrington_solvable_1975}
    	D.~Sherrington and S.~Kirkpatrick, ``Solvable {Model} of a {Spin}-{Glass},''
    	\emph{Physical Review Letters}, vol.~35, no.~26, pp. 1792--1796, Dec. 1975.
    	
    	\bibitem{camsari2017stochastic}
    	K.~Y. Camsari, R.~Faria, B.~M. Sutton, and S.~Datta, ``Stochastic p-bits for
    	invertible logic,'' \emph{Physical Review X}, vol.~7, no.~3, p. 031014, 2017.
    	
    	\bibitem{reichl1999modern}
    	L.~E. Reichl, ``A modern course in statistical physics,'' 1999.
    	
    	\bibitem{liu2015estimating}
    	Q.~Liu, J.~Peng, A.~Ihler, and J.~Fisher~III, ``Estimating the partition
    	function by discriminance sampling,'' in \emph{Proceedings of the
    		Thirty-First Conference on Uncertainty in Artificial Intelligence}.\hskip 1em
    	plus 0.5em minus 0.4em\relax AUAI Press, 2015, pp. 514--522.
    	
    	\bibitem{blackman_scrambled_2018}
    	D.~Blackman and S.~Vigna, ``Scrambled {Linear} {Pseudorandom} {Number}
    	{Generators},'' \emph{arXiv:1805.01407 [cs]}, May 2018, arXiv: 1805.01407.
    	
    	\bibitem{ortega-zamorano_high_2014}
    	F.~Ortega-Zamorano, J.~M. Jerez, G.~Ju{\'a}rez, J.~O. P{\'e}rez, and L.~Franco,
    	``High precision {FPGA} implementation of neural network activation
    	functions,'' in \emph{2014 {IEEE} {Symposium} on {Intelligent} {Embedded}
    		{Systems} ({IES})}, Dec. 2014, pp. 55--60.
    	
    	\bibitem{majzoobi_fpga_2010}
    	M.~Majzoobi, F.~Koushanfar, and S.~Devadas, ``{FPGA} {PUF} using programmable
    	delay lines,'' in \emph{2010 {IEEE} {International} {Workshop} on
    		{Information} {Forensics} and {Security}}, Dec. 2010, pp. 1--6.
    	
    	\bibitem{mahmoodi2019versatile}
    	M.~Mahmoodi, M.~Prezioso, and D.~Strukov, ``Versatile stochastic dot product
    	circuits based on nonvolatile memories for high performance neurocomputing
    	and neurooptimization,'' \emph{Nature communications}, vol.~10, no.~1, pp.
    	1--10, 2019.
    	
    	\bibitem{baity-jesi_janus_2014}
    	M.~Baity-Jesi, R.~A. Ba{\~n}os, A.~Cruz, L.~A. Fernandez, J.~M. Gil-Narvion,
    	A.~Gordillo-Guerrero, D.~I{\~n}iguez, A.~Maiorano, F.~Mantovani, E.~Marinari,
    	V.~Martin-Mayor, J.~Monforte-Garcia, A.~Mu{\~n}oz~Sudupe, D.~Navarro,
    	G.~Parisi, S.~Perez-Gaviro, M.~Pivanti, F.~Ricci-Tersenghi, J.~J.
    	Ruiz-Lorenzo, S.~F. Schifano, B.~Seoane, A.~Tarancon, R.~Tripiccione, and
    	D.~Yllanes, ``Janus {II}: {A} new generation application-driven computer for
    	spin-system simulations,'' \emph{Computer Physics Communications}, vol. 185,
    	no.~2, pp. 550--559, Feb. 2014.
    	
    	\bibitem{yoshimura_fpga-based_2016}
    	C.~Yoshimura, M.~Hayashi, T.~Okuyama, and M.~Yamaoka, ``{FPGA}-based
    	{Annealing} {Processor} for {Ising} {Model},'' in \emph{2016 {Fourth}
    		{International} {Symposium} on {Computing} and {Networking} ({CANDAR})}, Nov.
    	2016, pp. 436--442.
    	
    	\bibitem{ortega-zamorano_fpga_2016}
    	F.~Ortega-Zamorano, M.~A. Montemurro, S.~A. Cannas, J.~M. Jerez, and L.~Franco,
    	``{FPGA} {Hardware} {Acceleration} of {Monte} {Carlo} {Simulations} for the
    	{Ising} {Model},'' \emph{IEEE Transactions on Parallel and Distributed
    		Systems}, vol.~27, no.~9, pp. 2618--2627, Sep. 2016.
    	
    	\bibitem{suzuki1976relationship}
    	M.~Suzuki, ``Relationship between d-dimensional quantal spin systems and (d+
    	1)-dimensional ising systems: Equivalence, critical exponents and systematic
    	approximants of the partition function and spin correlations,''
    	\emph{Progress of theoretical physics}, vol.~56, no.~5, pp. 1454--1469, 1976.
    	
    	\bibitem{troyer2005computational}
    	M.~Troyer and U.-J. Wiese, ``Computational complexity and fundamental
    	limitations to fermionic quantum monte carlo simulations,'' \emph{Physical
    		review letters}, vol.~94, no.~17, p. 170201, 2005.
    	
    	\bibitem{santoro2002theory}
    	G.~E. Santoro, R.~Marto{\v{n}}{\'a}k, E.~Tosatti, and R.~Car, ``Theory of
    	quantum annealing of an ising spin glass,'' \emph{Science}, vol. 295, no.
    	5564, pp. 2427--2430, 2002.
    	
    	\bibitem{heim2015quantum}
    	B.~Heim, T.~F. R{\o}nnow, S.~V. Isakov, and M.~Troyer, ``Quantum versus
    	classical annealing of ising spin glasses,'' \emph{Science}, vol. 348, no.
    	6231, pp. 215--217, 2015.
    	
    	\bibitem{denchev2016computational}
    	V.~S. Denchev, S.~Boixo, S.~V. Isakov, N.~Ding, R.~Babbush, V.~Smelyanskiy,
    	J.~Martinis, and H.~Neven, ``What is the computational value of finite-range
    	tunneling?'' \emph{Physical Review X}, vol.~6, no.~3, p. 031015, 2016.
    	
    	\bibitem{hitachi_sqa_2017}
    	T.~Okuyama, M.~Hayashi, and M.~Yamaoka, ``An ising computer based on simulated
    	quantum annealing by path integral monte carlo method,'' in \emph{2017 IEEE
    		International Conference on Rebooting Computing (ICRC)}, Nov 2017, pp. 1--6.
    	
    	\bibitem{waidyasooriya_opencl-based_2019}
    	H.~M. Waidyasooriya, M.~Hariyama, M.~J. Miyama, and M.~Ohzeki,
    	``\BIBforeignlanguage{en}{{OpenCL}-based design of an {FPGA} accelerator for
    		quantum annealing simulation},'' \emph{\BIBforeignlanguage{en}{The Journal of
    			Supercomputing}}, Feb. 2019.
    	
    	\bibitem{camsari_scaled_2018}
    	K.~Y. Camsari, S.~Chowdhury, and S.~Datta, ``Scaled {Quantum} {Circuits}
    	{Emulated} with {Room} {Temperature} p-{Bits},'' \emph{arXiv:1810.07144
    		[quant-ph]}, Oct. 2018, arXiv: 1810.07144.
    	
    	\bibitem{pfeuty1970one}
    	P.~Pfeuty, ``The one-dimensional ising model with a transverse field,''
    	\emph{ANNALS of Physics}, vol.~57, no.~1, pp. 79--90, 1970.
    	
    	\bibitem{sachdev_2011}
    	S.~Sachdev, \emph{Quantum Phase Transitions}, 2nd~ed.\hskip 1em plus 0.5em
    	minus 0.4em\relax Cambridge University Press, 2011.
    	
    	\bibitem{hinton_coursera}
    	G.~Hinton, N.~Srivastava, and K.~Swersky, ``Neural networks for machine
    	learning,'' \emph{Coursera, video lectures}, no.~11, 2012.
    	
    	\bibitem{rizzo_fully_2013}
    	N.~D. Rizzo, D.~Houssameddine, J.~Janesky, R.~Whig, F.~B. Mancoff, M.~L.
    	Schneider, M.~DeHerrera, J.~J. Sun, K.~Nagel, S.~Deshpande, H.~Chia, S.~M.
    	Alam, T.~Andre, S.~Aggarwal, and J.~M. Slaughter, ``A {Fully} {Functional} 64
    	{Mb} {DDR}3 {ST}-{MRAM} {Builton} 90 nm {CMOS} {Technology},'' \emph{IEEE
    		Transactions on Magnetics}, vol.~49, no.~7, pp. 4441--4446, Jul. 2013.
    	
    	\bibitem{li_memristor-based_2013}
    	B.~Li, Y.~Shan, M.~Hu, Y.~Wang, Y.~Chen, and H.~Yang,
    	``\BIBforeignlanguage{en}{Memristor-based approximated computation},'' in
    	\emph{\BIBforeignlanguage{en}{International {Symposium} on {Low} {Power}
    			{Electronics} and {Design} ({ISLPED})}}.\hskip 1em plus 0.5em minus
    	0.4em\relax Beijing, China: IEEE, Sep. 2013, pp. 242--247.
    	
    	\bibitem{cai2019harnessing}
    	F.~Cai, S.~Kumar, T.~Van~Vaerenbergh, R.~Liu, C.~Li, S.~Yu, Q.~Xia, J.~J. Yang,
    	R.~Beausoleil, W.~Lu \emph{et~al.}, ``Harnessing intrinsic noise in memristor
    	hopfield neural networks for combinatorial optimization,'' \emph{arXiv
    		preprint arXiv:1903.11194}, 2019.
    	
    	\bibitem{kuon_measuring_2007}
    	I.~Kuon and J.~Rose, ``Measuring the {Gap} {Between} {FPGAs} and {ASICs},''
    	\emph{IEEE Transactions on Computer-Aided Design of Integrated Circuits and
    		Systems}, vol.~26, no.~2, pp. 203--215, Feb. 2007.
    	
    \end{thebibliography}

\end{document}